\pdfoutput=1
\documentclass[11pt,twoside,a4paper,cmspaper,final,collab]{cms-tdr}
\def\svnVersion{a8ba1e7}\def\svnDate{2026/07/15}\def\cmsCernNoTag{CERN-EP-2026-020}\def\cmsCernDate{\today}\def\cmsMessage{Published in Physical Review D as \href{http://dx.doi.org/10.1103/1b65-rblz}{\doi{10.1103/1b65-rblz}.}}
\begin{document}\cmsNoteHeader{B2G-24-022}

\newcommand{\pnt}{\ensuremath{\mathrm{PNT}_{\PQb\PQb}}\xspace}
\newcommand{\xbb}{\ensuremath{\PX\to\bbbar}\xspace}
\newcommand{\PY}{{\HepParticle{Y}{}{}}\Xspace}
\newcommand{\muF}{\ensuremath{\mu_{\text{F}}}\xspace}
\newcommand{\muR}{\ensuremath{\mu_{\text{R}}}\xspace}
\newcommand{\combine}{\textsc{Combine}\xspace}
\newcommand{\bbZZf}{\ensuremath{\Pb\Pb\PZ\PZ}\xspace}
\newcommand{\bbZZ}{\ensuremath{\Pb\Pb\PZ(\PQq\PQq,\PGn\PGn)\PZ(\Pell\Pell)}\xspace}
\newcommand{\bbZZqq}{\ensuremath{\Pb\Pb\PZ(\PQq\PQq)\PZ(\Pell\Pell)}\xspace}
\newcommand{\bbZZvv}{\ensuremath{\Pb\Pb\PZ(\PGn\PGn)\PZ(\Pell\Pell)}\xspace}
\newcommand{\bbWW}{\ensuremath{\Pb\Pb\PW\PW}\xspace}
\newcommand{\HH}{\ensuremath{\PH\PH}\xspace}
\newcommand{\HY}{\ensuremath{\PH\PY}\xspace}
\newcommand{\mll}{\ensuremath{m_{\Pell\Pell}}\xspace}
\newcommand{\mbb}{\ensuremath{m_{\PQb\PQb}}\xspace}
\newcommand{\HHbbZZ}{\ensuremath{\PH\PH\to \Pb\Pb\PZ\PZ}\xspace}
\newcommand{\zzqqll}{\ensuremath{\PZ\PZ\to \PQq\PQq\Pell\Pell}\xspace}
\newcommand{\zzqqllvv}{\ensuremath{\PZ\PZ\to \Pell\Pell\PQq\PQq\slash\PGn\PGn}\xspace}
\newcommand{\zzllvv}{\ensuremath{\PZ\PZ\to \Pell\Pell\PGn\PGn}\xspace}
\newcommand{\ppHH}{\ensuremath{\Pp\Pp\to \PX\to \HH}\xspace}
\newcommand{\ppHY}{\ensuremath{\Pp\Pp\to \PX\to \HY}\xspace}
\newcommand{\XtoHY}{\ensuremath{\PX\to \HY}\xspace}
\newcommand{\njet}{\ensuremath{n_{\text{jet}}}\xspace}
\newcommand{\Ni}{\ensuremath{N_{\mathrm{i}}}\xspace}
\newcommand{\Nii}{\ensuremath{N_{\mathrm{ii}}}\xspace}
\newcommand{\Niii}{\ensuremath{N_{\mathrm{iii}}}\xspace}
\newcommand{\mX}{\ensuremath{m^{\text{rec}}_{\PX}}\xspace}
\newcommand{\mXres}{\ensuremath{m_\PX}\xspace}
\newcommand{\mYres}{\ensuremath{m_\PY}\xspace}
\newcommand{\mXgen}{\ensuremath{m^{\text{gen}}_{\PX}}\xspace}
\newcommand{\mYgen}{\ensuremath{m^{\text{gen}}_{\PY}}\xspace}
\newcommand{\mXp}{\ensuremath{{m}^{\text{rec}}_{\PX^{\prime}}}\xspace}
\newcommand{\ppHYbbZZ}{\ensuremath{\Pp\Pp\to \PX\to \HY \to \Pb\Pb\PZ\PZ}\xspace}
\newcommand{\jhbb}{\ensuremath{\mathrm{J}^{\PH\PQb\PQb}}\xspace}
\newcommand{\bdtscore}{\ensuremath{s_{\text{BDT}}}\xspace}
\newcommand{\miniiso}{\ensuremath{\text{Iso}_{\text{mini}}}\xspace}

\newlength{\cmsFigWidth}
\ifthenelse{\boolean{cms@external}}{
  \providecommand{\cmsLeft}{upper\xspace}
  \providecommand{\cmsRight}{lower\xspace}
  \providecommand{\cmsTable}[1]{#1}
  \setlength{\cmsFigWidth}{\columnwidth}
  }{
    \providecommand{\cmsLeft}{left\xspace}
    \providecommand{\cmsRight}{right\xspace}
    \providecommand{\cmsTable}[1]{\resizebox{\textwidth}{!}{#1}}
    \setlength{\cmsFigWidth}{0.8\textwidth}
  }

\cmsNoteHeader{B2G-24-022}
\title{Search for a new resonance decaying to a Higgs boson and a scalar boson in events with two \texorpdfstring{\PQb}{b} jets and two \texorpdfstring{\PZ}{Z} bosons in proton-proton collisions at \texorpdfstring{$\sqrt{s}=13\TeV$}{sqrt(s) = 13 TeV}}

\author*[cern]{The CMS Collaboration}

\date{\today}

\abstract{
A search is performed for a new resonance \PX decaying into a Higgs boson and a new scalar boson \PY (\HY) including the case where \PY is assumed to be a second Higgs boson (\HH), using proton-proton collision data collected at $\sqrt{s}=13\TeV$, corresponding to an integrated luminosity of 138\fbinv. This study performs a comprehensive exploitation of the \bbZZf events, encompassing the following decay topologies. One \PH candidate is identified through its decay into a bottom quark-antiquark pair, while the other \PH or the \PY candidate is selected through its decay into a pair of \PZ bosons. 
One \PZ boson is required to decay leptonically and the other, to decay into a pair of quarks or neutrinos. Events of interest are categorized based on the Lorentz boosts of the hadronically decaying \PH and \PZ bosons. Machine-learning-based discriminants, together with the reconstructed resonance mass, are employed across the different categories to separate signal from backgrounds, and their corresponding distributions are included in a simultaneous fit. 
No significant deviations from the standard model predictions are observed. 
Upper limits at the 95\% confidence level are set on the \HY and \HH production cross sections. 
For \HY production, the upper limit on the cross section of the process \ppHYbbZZ ranges from 500 to 5\unit{fb}, with the most stringent constraints achieved for high-mass \PX resonances and \PY masses near the threshold for decay into a pair of massive vector bosons. This is comparable to the sensitivity achieved in other analyses, which focus on \PH decays to $\PGg\PGg$ or $\PGt\PGt$ and \PY decays into a pair of bottom quarks or massive vector bosons.
For resonant \HH production, the upper limit on the cross section of \ppHH production is 1\unit{pb} for a high-mass resonance.
}

\hypersetup{
pdfauthor={CMS Collaboration},
pdftitle={Search for a new resonance decaying to a Higgs boson and a scalar boson in events with two b jets and two Z bosons in proton-proton collisions at sqrt(s)=13 TeV},
pdfsubject={CMS},
pdfkeywords={CMS, BSM, resonant, di-Higgs, ZZ}}

\maketitle 

\section{Introduction}

The discovery of the Higgs boson (\PH) in 2012 by the ATLAS~\cite{2012tfa} and CMS~\cite{Chatrchyan:2012ufa,Chatrchyan:2012ufa_long} Collaborations at the CERN LHC marked the completion of the standard model (SM) of particle physics. 
Beyond the SM (BSM), numerous theoretical extensions predict the existence of heavy resonances (\PX) that decay into pairs of scalar bosons. Investigating the resonant production of a Higgs boson in association with a new scalar boson (\HY), as well as the Higgs boson pairs (\HH), serves as a powerful probe for exploring BSM physics. 
Studies of these processes typically search for a narrow-width resonance produced via an $s$-channel mediator, which is the approach taken in this analysis.

Supersymmetry (SUSY) is an extension of the SM that postulates a symmetry pairing fermions with bosons and vice versa~\cite{Golfand:1971iw, Wess:1974tw}. 
While the minimal supersymmetric standard model (MSSM) extends the SM Lagrangian by introducing another complex Higgs doublet~\cite{Fayet:1974pd, Fayet:1977yc}, the next-to-MSSM (NMSSM)~\cite{NMSSM2, NMSSM3} introduces one additional complex singlet field. 
The NMSSM is the most straightforward supersymmetric extension of the SM where the electroweak scale originates from the SUSY-breaking scale, providing a natural solution to the so-called $\mu$-problem of the MSSM~\cite{Kim:1983dt} by generating the $\mu$ parameter dynamically through the singlet's vacuum expectation value; this avoids the MSSM's need to manually set $\mu$ at the electroweak scale rather than its natural Planck-scale value.
In the NMSSM scenario, several spin-0 bosons emerge, one of which can be associated with the 125\GeV Higgs boson. 
The decay of the heavier scalar \PX may lead to signatures such as \XtoHY, where \PY denotes a new scalar boson. 
Similar decay topologies are also predicted by the two real scalar singlet model~\cite{trsm}. In this framework, for a high-mass \PY, the diboson decay modes $\PY \to \PZ\PZ/\PW\PW$ typically dominate the branching fractions, providing a direct experimental handle for investigating such extended scalar sectors.

In the limit where the parameters of these extended models yield identical scalars, or within distinct theoretical frameworks such as models with a warped extra dimension~\cite{PhysRevLett.83.4690,PhysRevLett.83.3370}, the symmetric $\PX \to \PH\PH$ decay mode becomes highly prominent. These extra-dimensional models postulate the existence of a small and compactified spatial dimension in which gravity carriers propagate. The ``Randall--Sundrum bulk model'' assumes that the matter particles are also allowed to propagate in the extra dimension. 
This model predicts the existence of massive resonances, including a spin-0 radion~\cite{PhysRevLett.83.4922,PhysRevD.62.045015,PhysRevD.63.065002} and a tower of Kaluza--Klein excitations of the graviton~\cite{PhysRevLett.84.2080,PhysRevD.62.046008,PhysRevD.76.036006}. Such resonances have significant branching fractions for decays into \HH.

\begin{figure}[h!]\centering
  \includegraphics[width=0.5\textwidth]{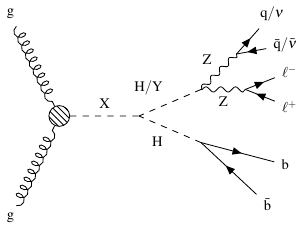}
  \caption{A diagram illustrating the gluon-gluon fusion production of a resonance \PX, which subsequently decays into either \HH or \HY. One of the Higgs bosons or the scalar \PY decays into two \PZ bosons. Of these, one \PZ boson decays either hadronically into a pair of quarks or invisibly into a pair of neutrinos, while the other \PZ boson decays leptonically into two charged leptons. The remaining Higgs boson decays into a pair of \PQb quarks.}
\label{fig:signal_feynman_diagram}
\end{figure}

In this analysis, we investigate the production of a resonance \PX decaying either into \HY or into \HH using proton-proton ($\Pp\Pp$) collision data recorded by the CMS experiment during Run 2 (2016--2018) at $\sqrt{s}=13\TeV$, corresponding to an integrated luminosity of 138\fbinv.
We consider masses of the resonance \PX between 700\GeV and 4\TeV for \HY decays, with the mass of the resonance \PY chosen to ensure the SM Higgs boson remains on shell.
For \HH decays, masses of the resonance \PX between 260\GeV and 4\TeV are studied.
The decay mode of \HY (\HH) considered is as follows: one of the Higgs bosons decays into a pair of bottom (\Pb) quarks, while the scalar \PY (or the other Higgs boson) decays into two \PZ bosons. Of the two \PZ bosons, one decays into a pair of charged leptons (\Pe, \Pgm), including cases where the \Pe or \Pgm results from the decay of a \Pgt. The other \PZ boson decays either hadronically into two quarks or invisibly into two neutrinos (\Pgne, \Pgngm, \Pgngt). In summary, the final state is labeled as {\bbZZ}, where $\Pell$ corresponds to an electron or a muon (referred to as the electron or muon channel, respectively). 
Figure~\ref{fig:signal_feynman_diagram} illustrates a representative diagram for the process under study.
Additionally, the \bbWW process arising from the decay of resonance \PX, with both \PW bosons decaying leptonically, has the same final state as \bbZZvv, and it can also contribute in the \bbZZqq channel, as additional jets can originate from quarks or gluons from initial-state radiation (ISR) or final-state radiation. 

Compared to a similar CMS search using 2016 data focusing on \HH production~\cite{PhysRevD.102.032003}, the present analysis shifts its primary focus to the \HY production process, which remains unexplored in this specific channel, while simultaneously updating the \HH search as a natural extension. 
To achieve optimal sensitivity across these distinct topologies, several key improvements have been incorporated. First, a more restrictive selection is employed to ensure orthogonality to a dedicated \bbWW analysis~\cite{2024bbww}. 
Such a strategy is pivotal to establishing a clear and unambiguous definition of the signal process, effectively preventing signal contamination from the overlapping \bbWW process. 
This treatment is essential due to the similar final-state signatures of the two processes.
With this restrictive selection, \bbWW makes a minor contribution in the \bbZZqq channel, however it still accounts for up to 20\% of the signal region in the \bbZZvv channel. Thus, it is explicitly included as part of the signal in the analysis. 
Furthermore, the search is extended to a broader mass spectrum by incorporating additional measurement regions, such as those targeting Lorentz-boosted objects, which enhances the sensitivity to high-mass resonances. 
The results obtained for the \HY channel provide essential complementarity to alternative searches focusing on \PH decays to $\PGg\PGg$ or $\Pgt\Pgt$ and \PY decays into a pair of bottom quarks or massive vector bosons~\cite{2025368,2023atlasvvtata,atlasbbgg}, together placing stringent model-independent constraints on the allowed BSM parameter space.

\section{The CMS detector and simulated events}
The CMS apparatus~\cite{Chatrchyan:2008zzk,CMS:2023gfb} is a multipurpose, nearly hermetic detector, designed to trigger on~\cite{CMS:2020cmk,CMS:2016ngn,CMS:2024aqx} and identify electrons, muons, photons, and (charged and neutral) hadrons~\cite{CMS:2020uim,CMS:2018rym,CMS:2014pgm}. Its central feature is a superconducting solenoid of 6\unit{m} internal diameter, providing a magnetic field of 3.8\unit{T}. Within the solenoid volume are a silicon pixel and strip tracker, a lead tungstate crystal electromagnetic calorimeter (ECAL), and a brass and scintillator hadron calorimeter (HCAL), each composed of a barrel and two endcap sections. Forward calorimeters extend the pseudorapidity coverage provided by the barrel and endcap detectors. Muons are reconstructed using gas-ionization detectors interleaved with the layers of the steel flux-return yoke outside the solenoid.

Events of interest are selected using a two-tiered trigger system. The first level, composed of custom hardware processors, uses information from the calorimeters and muon detectors to select events at a rate of around 100\unit{kHz} within a fixed latency of 4\mus~\cite{CMS:2020cmk}. The second level, known as the high-level trigger, consists of a farm of processors running a version of the full event reconstruction software optimized for fast processing and reduces the event rate to a few kilohertz before data storage~\cite{CMS:2016ngn,CMS:2024aqx}. A more detailed description of the CMS detector, together with a definition of the coordinate system and kinematic variables, can be found in Refs.~\cite{Chatrchyan:2008zzk,CMS:2023gfb}.

The simulated signal events are produced at leading order (LO) using \MGvATNLO 2.6.5~\cite{Alwall:2014hca} (hereafter referred to as MG) to resonance decay level, \ie, $\Pp\Pp\to \PX\to \PH\PY$ or $\Pp\Pp\to\PX\to \PH\PH$. The decay of \PY/\PH and the subsequent \PZ boson decay are handled by \PYTHIA~8.240~\cite{Sjostrand:2014zea}.

The primary backgrounds are Drell--Yan (DY) produced lepton pairs and fully leptonic decays of top quark-antiquark pairs (hereafter referred to as \ttbar). The DY samples are generated at next-to-leading order (NLO) in quantum chromodynamics (QCD) using MG, with up to two additional partons included in the matrix element calculations. The ``FxFx'' merging method~\cite{Frederix_2012} is used to handle the matching between the jets from matrix element calculations and parton showering to avoid double counting. 
The event yield of DY production is scaled to the cross section computed at next-to-next-to-leading order (NNLO) precision in perturbative QCD, including NLO electroweak corrections calculated by \FEWZ 3.1~\cite{Gavin:2010az}. The \ttbar sample is generated at NLO using \POWHEG v2~\cite{Frixione_2007}, and its production cross section is rescaled to the NNLO value~\cite{Czakon:2011xx}.

The diboson, triboson, {\ttbar}+\PV (\PV = \PZ,\PW), and top+\PV samples are generated at NLO using either \POWHEG or MG. The {\ttbar}+$\PV\PV$ samples are generated at LO using MG. The SM Higgs boson production (\eg, via vector boson or gluon-gluon fusion) is generated at NLO using \POWHEG.

Processes with fewer than two leptons produced in the hard interaction (prompt leptons), \eg, \ttbar ($\bbbar\Pell\nu\PQq\PQq$), {\PW}+jets, and QCD multijets are simulated using \POWHEG, MG, and \PYTHIA, respectively.
They can emulate the signal through nonprompt or misidentified leptons and their contributions are estimated using data in signal-depleted control regions (CRs). Those simulated samples are used to assess the systematic uncertainties of the nonprompt-lepton backgrounds.

We use the \PYTHIA generator package for parton showering, fragmentation, and hadronization. The \PYTHIA parameters that affect the description of the underlying event are set according to the CP5 tune~\cite{CP5}. The NNPDF 3.1~\cite{nnpdf31} parton distribution functions (PDFs) at NNLO are used in these studies, and the CMS detector response in simulated events is modeled using the \GEANTfour package~\cite{geant4, geant4_2}. 
We use a ``tag-and-probe'' procedure~\cite{tagandprobe} to measure factors to correct for differences in efficiencies between data and simulation in the trigger, reconstruction, and selection. The simulated events include additional $\Pp\Pp$ interactions in the same and neighboring bunch crossings, referred to as pileup. Simulated events are weighted so the pileup distribution matches the one from data.

\section{Event selection}
Events of interest comprise a dilepton pair ($\Pe\Pe$ or $\Pgm\Pgm$) and jets, supplemented by neutrinos specifically within the {\bbZZvv} channel.
Event selection is based on a combination of triggers requiring the presence of either one or two electrons or muons.
The average offline trigger efficiencies exceed 95\% for both electron and muon channel events, reaching up to 98\% as the transverse momentum (\pt) of the leptons increase, specifically for electrons with $\pt>100\GeV$ and muons with $\pt>60\GeV$. 

\subsection{Object reconstruction}
A particle-flow (PF) algorithm~\cite{CMS-PAPERS-PRF-14-001} aims to reconstruct and identify each individual particle in an event, with an optimized combination of information from the various elements of the CMS detector. 
The energy of photons is obtained from the ECAL measurement. 
The energy of an electron is determined from a combination of the electron momentum at the primary interaction vertex as determined by the tracker, the energy of the corresponding ECAL cluster, and the energy sum of all bremsstrahlung photons spatially compatible with originating from the electron track. 
The energy of a muon is obtained from the curvature of the corresponding track. 
The energy of a charged hadron is determined from a combination of the momentum measured in the tracker and the matching ECAL and HCAL energy deposits, corrected for the response function of the calorimeters to hadronic showers. 
Finally, the energy of neutral hadrons is obtained from the corresponding corrected ECAL and HCAL energies. 

The primary vertex (PV) is taken to be the vertex corresponding to the hardest scattering in the event, evaluated using tracking information alone, as described in Section 9.4.1 of Ref.~\cite{CMS-TDR-15-02}.
We require each event to have at least one good PV, defined as having a number of degrees of freedom (ndof) ${\geq}4$, where ndof reflects the weighted contribution of tracks used in the vertex reconstruction. In addition, the PV must lie in the central detector region of $\abs{z} < 24$ and $\rho < 2\unit{cm}$ around the nominal interaction point.

The reconstruction of electron objects involves combining the information from clusters of energy deposited in the ECAL and the trajectories fitted in the tracker~\cite{Khachatryan:2015hwa}.
They are required to have $\pt>10\GeV$ and $\abs{\eta}<2.5$, while the ECAL transition region $1.44<\abs{\eta}<1.57$ is excluded.
Multivariate analysis (MVA)~\cite{Hocker:2007ht} identification criteria based on electron kinematic properties suppress misidentification~\cite{CMS:2020uim}; the MVA omits isolation variables because the electrons can be very close to the jets from \PZ boson decay.
Here, the isolation variable is calculated by summing the \pt of all charged hadrons and neutral particles found within a predefined geometrical cone.
Two MVA working points (WPs) are used in the analysis, the ``loose'' WP (WPL) with efficiency of 98\% and the ``tight'' WP (WPT) with efficiency of 90\%. The WPL requirement defines the loose electron.

The tight electron is defined using the WPT with additional requirements on the PV, \ie, impact parameters $\abs{d_{xy}}$ and $\abs{d_{z}}$, the three-dimensional impact parameter significance,
the relative mini-isolation (hereafter referred to as {\miniiso}), and the requirement $\pt>15\GeV$. 
Mini-isolation is similar to isolation; 
however, the cone size $\Delta R = \sqrt{\smash[b]{(\Delta\eta)^2+(\Delta\phi)^2}}$  of mini-isolation changes with respect to the \pt of lepton ${\pt}_{\text{,lep}}$~\cite{miniiso2018} in the following way: 
\begin{equation}
    \Delta R =
    \begin{cases}
    0.2, & {\pt}_{\text{,lep}} < 50\GeV,\\
    \frac{10\GeV}{{\pt}_{,\text{lep}}}, & {\pt}_{\text{,lep}} \in (50, 200)\GeV, \\
      0.05, & {\pt}_{\text{,lep}} > 200\GeV.
    \end{cases}
    \label{equ:miniiso}
\end{equation}
The {\miniiso} is defined as the ratio of the mini-isolation variable to the \pt of the lepton. We require the tight electrons to satisfy {\miniiso} smaller than 0.2.

Muons are reconstructed from information in the muon system and the inner tracker with $\abs{\eta}<2.4$ and $\pt>10\GeV$~\cite{CMS:2018rym}. 
Muon candidates must satisfy identification criteria based on the number of hits in the muon system and the inner tracker, the quality of the combined fit to a track, the number of matched muon detector planes, and the consistency between the track and the PV (hereafter referred to as threshold-based ID). 
Similar to the case of electrons, two types of muons are defined. The threshold-based WPL is used for both loose and tight muons; the tight muon is defined with additional requirements on track quality, muon segment quality, and an {\miniiso} of less than 0.2.

For both the electron and muon channels, nonisolated leptons are defined. They satisfy the same selection criteria as a tight lepton, except that the {\miniiso} is required to be within the range [0.2, 0.4]. 
We use these nonisolated leptons to estimate the nonprompt-lepton background.

Jets are reconstructed using PF objects and the anti-\kt jet clustering algorithm~\cite{Cacciari:2008gp,Cacciari:2011ma} as implemented in the \textsc{FastJet} package~\cite{Cacciari:2011ma}.
Jets reconstructed excluding charged particles associated with pileup vertices, using a technique known as charged-hadron subtraction (CHS)~\cite{CMS-PAPERS-PRF-14-001}, are referred to as CHS jets. 
The pileup-per-particle identification (PUPPI) algorithm~\cite{Sirunyan:2020foa,Bertolini:2014bba} is used to mitigate the effect of pileup at the reconstructed particle level, making use of local shape information, event pileup properties, and tracking information.
A local shape variable distinguishes between collinear and soft diffuse distributions of other particles surrounding the particle under consideration.
The former is attributed to particles originating from the hard scatter and the latter to particles originating from pileup interactions. Charged particles identified to be originating from pileup vertices are discarded. 
For each neutral particle, a local shape variable is computed using the surrounding charged particles compatible with the PV within the tracker acceptance ($\abs{\eta}<2.5$), and using both charged and neutral particles in the region outside of the tracker coverage. 
The momenta of the neutral particles are then rescaled according to the probability that they originate from the primary interaction vertex deduced from the local shape variable, superseding the need for jet-based pileup corrections~\cite{Sirunyan:2020foa}. Jets reconstructed with the PUPPI algorithm are referred to as PUPPI jets.

Two types of jets are defined to target the resolved and boosted decays of \PH and \PZ bosons. Large-radius jets, reconstructed with a distance parameter of 0.8, are referred to as AK8 jets, while AK4 jets are reconstructed with a distance parameter of 0.4. The PUPPI algorithm is used for the reconstruction of AK8 jets, whereas the CHS algorithm is applied for AK4 jets.
The AK8 jets are required to have $\pt > 300\GeV$ and a soft-drop mass (${m}_{\mathrm{SD}}$)~\cite{Larkoski_2014} greater than 30\GeV. Additionally, all the AK8 jets are required to be separated by $\Delta R>0.8$ from all the AK4 jets. 
The AK4 jets must have $\pt > 30\GeV$. Both AK8 and AK4 jets are required to lie within a pseudorapidity range $\abs{\eta}<2.4$.

To address the overlap between leptons and jets---which commonly would occur between leptons and jets from the two \PZ bosons produced in the decay of a very heavy resonance---all jet candidates undergo lepton subtraction. In this process, the four-momentum of the lepton is subtracted from the four-momentum of the jet if the spatial separation $\Delta R$ is smaller than 0.8 for AK8 and 0.4 for AK4 jets. The raw \pt and energy of the jet are used during the lepton subtraction.

The AK4 jets originating from the \PQb quarks are identified (\PQb-tagged) by \textsc{DeepJet}~\cite{Bols_2020,CMS-DP-2018-058}: a deep neural network algorithm that takes as input the tracks within the jet that are displaced from the PV, 
the identified secondary vertices, the jet kinematic variables, and the neutral PF candidates. 
We use a threshold on the  $\textsc{DeepJet}$ score that corresponds to a 10\% probability of misidentifying a light-flavor quark- or a gluon-induced AK4 jet as a \PQb quark jet and maintains a 87\% \PQb tagging efficiency. Additionally, we require all the \PQb-tagged AK4 jets to have a separation $\Delta R>0.4$ from all the tight leptons.

The AK8 jets originating from Lorentz-boosted Higgs boson decays are identified using the ParticleNet (PNet) algorithm~\cite{PhysRevD.101.056019,Hayrapetyan_2025}. The \textsc{PNet} algorithm has an advanced architecture based on a dynamic graph convolutional neural network, which exploits the jet constituents to classify jets originating from different types of quarks.
This work uses the mass-decorrelated (MD)~\cite{Dolen:2016kst} version of this algorithm. This version ensures that the mass distributions of nonresonant backgrounds are not altered by a selection based on the ParticleNet-MD algorithm score~\cite{CMS:2020poo}.

The discriminant used to distinguish $\PH \to \bbbar$ AK8 jets from QCD-induced jets is defined as a binary classification score~\cite{Hayrapetyan_2025} (referred to as ``\pnt''):
\begin{equation}
    \mathrm{PNT}_{\PQb\PQb}=\frac{P(\PX \to \bbbar)}{P(\PX \to \bbbar) + P(\mathrm{QCD})},
\end{equation}
where $P(\PX\to \bbbar)$ and $P(\mathrm{QCD})$ can be interpreted as the probabilities that a jet originates from a spin-0 particle \xbb decay and from QCD-induced processes, respectively. 
We use a threshold on the \pnt corresponding to 80\% signal efficiency, where the probability of misidentifying a QCD-induced AK8 jet as a \xbb AK8 jet is a few percent. 
The AK8 jet that satisfies this threshold is referred to as \jhbb, and it must be spatially separated from tight leptons by a distance $\Delta R>0.8$. 

Jet energy corrections are derived from simulation studies so that the average measured energy of the corrected jets is identical to that of jets at the particle level. $\textit{In situ}$ measurements of the momentum balance in dijet, $\gamma$+jet, {\PZ}+jet, and multijet events are used to determine any residual differences between the jet energy scale in data and simulation, and appropriate corrections are made~\cite{jer}. Additional selection criteria are applied to each jet to remove jets originating from instrumental effects or reconstruction failures~\cite{Sirunyan:2020foa}.

The missing transverse momentum vector \ptvecmiss is computed as the negative vector sum of the transverse momenta of all the PF candidates in an event, and its magnitude is denoted as \ptmiss~\cite{CMS:2019ctu}.

\subsection{Event categorization}

Given the wide range considered in the mass of the resonance \PX, we define four signal regions (SRs) to capture the topologies of the decay products from the \PH and \PZ bosons in the \bbZZqq signals.
The common selection criteria for these four SRs are as follows. Each event must have exactly two tight leptons (electrons or muons, referred to as $n^{\text{tight}_{\Pell}}$) with opposite electric charge and same flavor. 
For the electron channel, the \pt of the leading lepton must be greater than 25\GeV and that of the subleading lepton greater than 15\GeV.
For the muon channel, the \pt of the leading lepton must exceed 20\GeV and that of the subleading lepton must exceed 12\GeV.
The number of additional loose leptons ($n^{\text{loose}_{\Pell}}$) in each event is required to be zero.
The invariant mass of the dilepton pair (\mll) must be within the range 75--105\GeV to suppress the contribution from {\bbWW}.
The \pt of the dilepton system ($\pt^{\Pell\Pell}$) must be greater than 10\GeV. The \ptmiss must be smaller than 100\GeV in order to ensure that the parameter space of {\bbZZqq} remains orthogonal to those of the {\bbZZvv} signals. The following jet-related requirements define the four SRs (where ``M'' stands for ``merged jet''):

\begin{itemize}
\item The SR ``qq0M,'' which corresponds to the signatures of the resolved decay of both the \PH and \PZ bosons, requires the absence of AK8 jets and the presence of at least four AK4 jets. Among these AK4 jets, at least two must be \PQb-tagged. 
If more than two \PQb-tagged jets are present, the two with invariant mass closest to the SM Higgs boson mass are selected. The invariant mass of the two selected \PQb-tagged jets (\mbb) must lie within the range [90, 150]\GeV. 
These two \PQb-tagged jets are assigned to the Higgs boson decay. The two leading jets among the remaining AK4 jets are then assigned to \PZ boson decay.
\item The SR ``qqHM,'' which targets events with a boosted Higgs boson decay and a resolved \PZ boson decay, requires exactly one AK8 jet that passes the \pnt threshold, as well as at least two AK4 jets. 
The AK8 jet is assigned to the Higgs boson decay and the two leading AK4 jets are assigned to the \PZ boson decay.
\item The SR ``qqZM,'' which targets events with a resolved Higgs boson decay and merged jets from the \PZ boson decay, requires exactly one AK8 jet which fails the \pnt threshold, along with at least two \PQb-tagged AK4 jets. 
If more than two \PQb-tagged jets are present in an event, the two with invariant mass closest to the SM Higgs boson mass are selected. 
These two \PQb-tagged AK4 jets are assigned to the Higgs boson decay and the AK8 jet is assigned to the \PZ boson decay.
\item The SR ``qq2M,'' which targets events with merged jets from decays of both the \PH and \PZ bosons, requires at least two AK8 jets, with at least one satisfying the \pnt threshold. The AK8 jet with the higher \pnt is assigned to the Higgs boson decay, and the leading AK8 jet among the remaining AK8 jets is assigned to \PZ boson decay.
\end{itemize}

It should be noted that no jet substructure requirements are applied to the AK8 jets originating from hadronic \PZ boson decays. This is because the internal jet energy distribution is significantly altered during the lepton subtraction, meaning that standard jet tagging performance and its associated calibrations are no longer applicable.

Similarly, we categorize the \bbZZvv events into two SRs according to the decay topology of the Higgs boson. The selection criteria on the leptons are the same as those in the \bbZZqq channel. The \ptmiss must be greater than 100\GeV. The following requirements, all related to the jets, define the two SRs:
\begin{itemize}
\item The SR ``$\PGn\PGn$1M,'' which targets events with merged jets from the \PH boson decay, requires at least one AK8 jet that satisfies the {\pnt} threshold.
\item The SR ``$\PGn\PGn$0M,'' which targets events with a resolved \PH boson decay, requires at least two \PQb-tagged AK4 jets. If more than two \PQb-tagged jets are present in an event, the two with invariant mass closest to the Higgs boson mass are selected.
\end{itemize}

In addition, we define CRs to study the dominant backgrounds, DY and \ttbar. Because the heavy-flavor component of the DY background is known to be challenging to model, two DY CRs are defined to closely match the flavor composition of the SRs.
The first DY CR (DYCR1) is defined similar to the SR qq0M, with the exception that it requires the presence of at least two \PQb-tagged AK4 jets and imposes a condition: $m_{\PQb\PQb}<90$ or $m_{\PQb\PQb}>150\GeV$. 
We use DYCR1 to extract the DY background normalization factor, which we then apply to DY events in SRs qq0M, qqZM, and $\PGn\PGn$0M, where the flavor compositions are similar to that of DYCR1.
The second DY CR (DYCR2) shares the same lepton selection criteria as the SR qqHM. It requires the presence of at least one AK8 jet and at least two AK4 jets. Additionally, we require that no AK8 jet passes the {\pnt} threshold, and that no more than one \PQb-tagged AK4 jet is present. 
The DYCR2 is designed to extract the DY normalization factor in SR qqHM, qq2M, and $\PGn\PGn$1M, while remaining orthogonal to all SRs and to DYCR1.

The \ttbar CR (TTCR) shares the same requirements as the SRs for lepton \pt and lepton $\eta$. However, it imposes an additional condition that the dilepton mass must be greater than 105\GeV. Furthermore, the event must contain at least two \PQb-tagged AK4 jets, and the \ptmiss is required to exceed 50\GeV.

The requirements for the SRs and CRs are summarized in Table~\ref{tab:SRCR_def}.

\begin{table*}[ht]
    \centering
    \topcaption{Definitions of all SRs and CRs. Here, $\pt^{\Pell_1}$ ($\pt^{\Pell_2}$) denotes the transverse momentum of the leading (subleading) lepton; ${n}^\text{AK8 jet}$, ${n}^{\PH\PQb\PQb}$, ${n}^\text{AK4 jet}$, and ${n}^\text{AK4 \PQb jet}$ represent the number of AK8 jets, AK8 jets passing the {\pnt} threshold, AK4 jets, and \PQb-tagged AK4 jets, respectively. All energy-related quantities are given in \GeV. The symbol ``\NA'' indicates that no cut is applied.}
    \renewcommand{\arraystretch}{1.3}
    \cmsTable{
      \begin{scotch}{lccccccccc}
          & qq2M & qqHM & qqZM & qq0M & $\PGn\PGn$1M & $\PGn\PGn$0M & DYCR1 & DYCR2 & TTCR \\
          \hline
          $n^{\text{tight}_{\Pell}} $ & \multicolumn{9}{c}{2} \\
          $n^{\text{loose}_{\Pell}} $ & \multicolumn{9}{c}{0} \\
          $\pt^{\Pell_1}$ &  \multicolumn{9}{c}{25 (20) for electron (muon) channel events} \\
          $\pt^{\Pell_2}$ &  \multicolumn{9}{c}{15 (12) for electron (muon) channel events } \\
          $m_{\Pell\Pell}$ & \multicolumn{8}{c}{$75\leq m_{\Pell\Pell} \leq105$} & ${>}105$\\
          $\pt^{\Pell\Pell}$ & \multicolumn{8}{c}{${>}10$} & \NA \\
          $n^\text{AK8 jet}$ & ${\geq}2$ & =1 & =1 & =0 & ${\geq}1$ & =0 & =0 & ${\geq}1$ & \NA \\
          $n^{\PH\PQb\PQb}$ & ${\geq}1$ & =1 & =0 & =0 & ${\geq}1$ & =0 & =0 & =0 & \NA\\
          $n^\text{AK4 jet}$ & \NA & ${\geq}2$ & ${\geq}2$ & ${\geq}4$ & \NA & ${\geq}2$ & ${\geq}2$ & ${\geq}2$ & ${\geq}2$ \\
          $n^\text{AK4 \PQb jet}$ & \NA & \NA & ${\geq}2$ & ${\geq}2$ & \NA & ${\geq}2$ & ${\geq}2$ & ${<}2$ & ${\geq}2$ \\
          \mbb & \NA & \NA & \NA & [90, 150] & \NA & [90, 150] & ${>}150$ or ${<}90$ & \NA & \NA \\ 
          \ptmiss & ${<}100$ & ${<}100$ & ${<}100$ & ${<}100$ & ${>}100$ & ${>}100$ & ${<}100$ & ${<}100$ & ${>}50$ 
      \end{scotch}
    }
    \label{tab:SRCR_def}
\end{table*}

\section{Background estimation and signal extraction strategy}
\subsection{\texorpdfstring{Estimation of DY and \ttbar processes}{Estimation of DY and ttbar processes}}
The DY and \ttbar processes are the two dominant backgrounds. In DY events, \PQb quarks can arise from ISR or gluon splitting in the final state. 
Additionally, light-flavor quarks can be misidentified as \PQb quarks during jet identification. Although the probability of such scenarios is small, their contribution cannot be neglected because of the very large cross section of the DY process.

Modeling jet multiplicity (\njet, referring to AK4 jets) in the DY process is known to be highly challenging, especially in the presence of heavy-flavor jets. Since the SRs require between two and four jets, an accurate estimation of the DY contribution across different \njet bins is essential. 
To determine the normalization factors, we fit to the \mll distribution separately for events with $\njet=2$, $\njet=3$, and $\njet\geq4$. 
The DY background shape is modeled as a convolution of the simulated \mll distribution with a Gaussian function to account for the resolution difference between data and simulation. 
Similarly, the \mll distributions of all other background processes are summed and convolved with a Gaussian function to form the total background shape.
We then fit the \mll distribution in data, where the background normalization is fixed, and only the DY normalization is allowed to vary. 
The normalization factor is defined as the ratio of the DY yields obtained from fit to that from simulation. 
Its uncertainty is dominated by the statistical uncertainty of the fit.
No additional extrapolation uncertainty is assigned, as the relative contributions of different jet flavors in DY events are found to be consistent between the CRs and the corresponding SRs.
The fit is performed separately in the DYCR1 and the DYCR2 and in the electron and muon channels.

The \ttbar process produces two prompt leptons and two \PQb quarks from the hard scattering, resulting in a topology that closely resembles that of the SRs. As a result, it is the dominant background in the SR $\PGn\PGn$0M and a second-most-important contribution in the other SRs. 
The \ttbar normalization factor is extracted in the TTCR, where the \ttbar contribution is dominant. 
Ratios of various kinematic distributions, such as \mll, \njet, and $\pt^{\Pell\Pell}$, between data and \ttbar simulation are found to be flat, justifying an inclusive extraction of the normalization factor.
It is defined as the ratio of the data yield (after subtracting all non-\ttbar background contributions) to the simulated \ttbar yield. We obtain the associated uncertainty by propagating the uncertainties from both the data and simulated events yields.

\subsection{Background estimation of nonprompt-lepton process}
The nonprompt-lepton process represents another background source. Nonprompt leptons arise from various mechanisms, including heavy-flavor hadron decays, misidentified hadrons, muons from decays of light mesons, or photon conversions within jets. 
The primary contributor to nonprompt leptons is the \ttbar process in the $\Pell$+jets channel. Additional contributions come from processes such as {\PW}+jets, single top quark, and QCD.

We estimate the contribution from nonprompt-lepton processes using a method based on control samples in data. We define three sideband regions by inverting the lepton selection criteria of the SRs. 
Specifically, these regions are defined as requiring (i) two leptons with same-sign (SS) charge; (ii) two SS leptons where at least one lepton is nonisolated; and (iii) two opposite-sign leptons where at least one lepton is nonisolated.
We estimate the yields of nonprompt-lepton backgrounds in the SR  using the relation $N_{\mathrm{SR}}=\Niii (\Ni/\Nii)$, where \Ni, \Nii, and \Niii represent the nonprompt yields in the three sidebands.

The yields of nonprompt processes in the first two sideband regions, \ie, \Ni and \Nii, are obtained by subtracting the contributions from all processes with at least two prompt leptons from the data yields. 
In contrast, we take the yields in the third sideband region, \Niii, directly from data events. The nonprompt-lepton background estimation is performed separately for the electron and muon channels.
As a cross-check, nonprompt yields in the SRs are also estimated using simulated samples. The difference between the yields obtained from the two methods is taken as the dominant uncertainty in the nonprompt-lepton background estimate, amounting to about 20\% in both electron and muon channels.

\subsection{Signal extraction strategy}
We estimate contributions from all other processes involving at least two prompt leptons, such as multiboson production, \ttbar production with associated vector bosons, and SM Higgs boson processes, through simulations. These estimates are normalized using either the most recent measurement results or state-of-the-art high-order theoretical calculations to ensure precision~\cite{CMS:2017zmo,LHCHiggsCrossSectionWorkingGroup:2016ypw,Frederix:2021agh,Maltoni:2015ena}.

Given the presence of a BSM resonance in the signal, the resonance mass serves as a straightforward and powerful variable for distinguishing signal from background events. 
In the case of \bbZZqq signal events, the event can be fully reconstructed regardless of the decay topology.
Specifically, the \PY scalar or the Higgs boson can be reconstructed from the two \PZ bosons via their decay products, \ie, two leptons and two AK4 jets or one AK8 jet. The other Higgs boson is reconstructed from either two \PQb-tagged AK4 jets or the AK8 jet identified as \jhbb.
In this case, the resonance mass \mX is reconstructed from the Higgs boson and the scalar \PY or from the two Higgs bosons.

In the case of \bbZZvv signal events, however, the event can no longer be fully reconstructed with visible objects because of the presence of neutrinos in the final state. 
To address this, the four-momentum of the \PZ boson decaying to neutrinos is replaced by a Lorentz vector with $E=\ptmiss$, $\ptvec=\ptvecmiss$, and $p_z=0$. A transverse resonance mass \mXp can then be reconstructed using the same expression that was used to define the \mX in the fully reconstructed \bbZZqq case,
\begin{equation}
    {m}^\text{rec}_{\PX^{\prime}}=\sqrt{m_{\PH\PZ}^2 + 2(\ptmiss E_{\PH\PZ} - \ptvecmiss \cdot \vec{p}_{\PH\PZ})},
    \label{equ:mxp}
\end{equation}
where $m_{\PH\PZ}$, $E_{\PH\PZ}$, and $\vec{p}_{\PH\PZ}$ denote the mass, energy, and momentum of the system composed of the hadronically decayed Higgs boson and the leptonically decayed \PZ boson.

Among the six SRs, the two fully resolved decay SRs, \ie, qq0M and $\PGn\PGn$0M, employ a boosted decision tree (BDT)~\cite{bdt} to distinguish signal events from background events. 
For the remaining SRs, the resonance mass distribution (\mX or \mXp) is used to extract the signal strength, as the limited signal acceptance results in insufficient background events for effective BDT training.

In this analysis, 383 mass points are scanned for the \HY signal and 35 mass points are scanned for the \HH signal. Given the large number of signal scenarios considered, we employ a BDT that incorporates the resonance mass as a training feature, trained separately for each data-taking year and lepton channel (electron and muon).
For signal events, the resonance mass corresponds to the nominal value from the simulation (\ie, \mXgen for mass of resonance \PX and \mYgen for mass of resonance \PY), which is assigned to all Monte Carlo (MC) events within a given sample.
For background processes, a mass value is randomly assigned to each event from a discrete uniform distribution over the set of considered mass points.
During training, signal events from different samples are assigned equal weights.
Specifically, for the \HY search, a separate BDT is trained for each \mXgen, utilizing \mYgen as an additional input feature. For the \HH search, a single BDT is trained using \mXgen as an input.

\begin{figure*}[hbtp]\centering
  \includegraphics[width=0.48\textwidth]{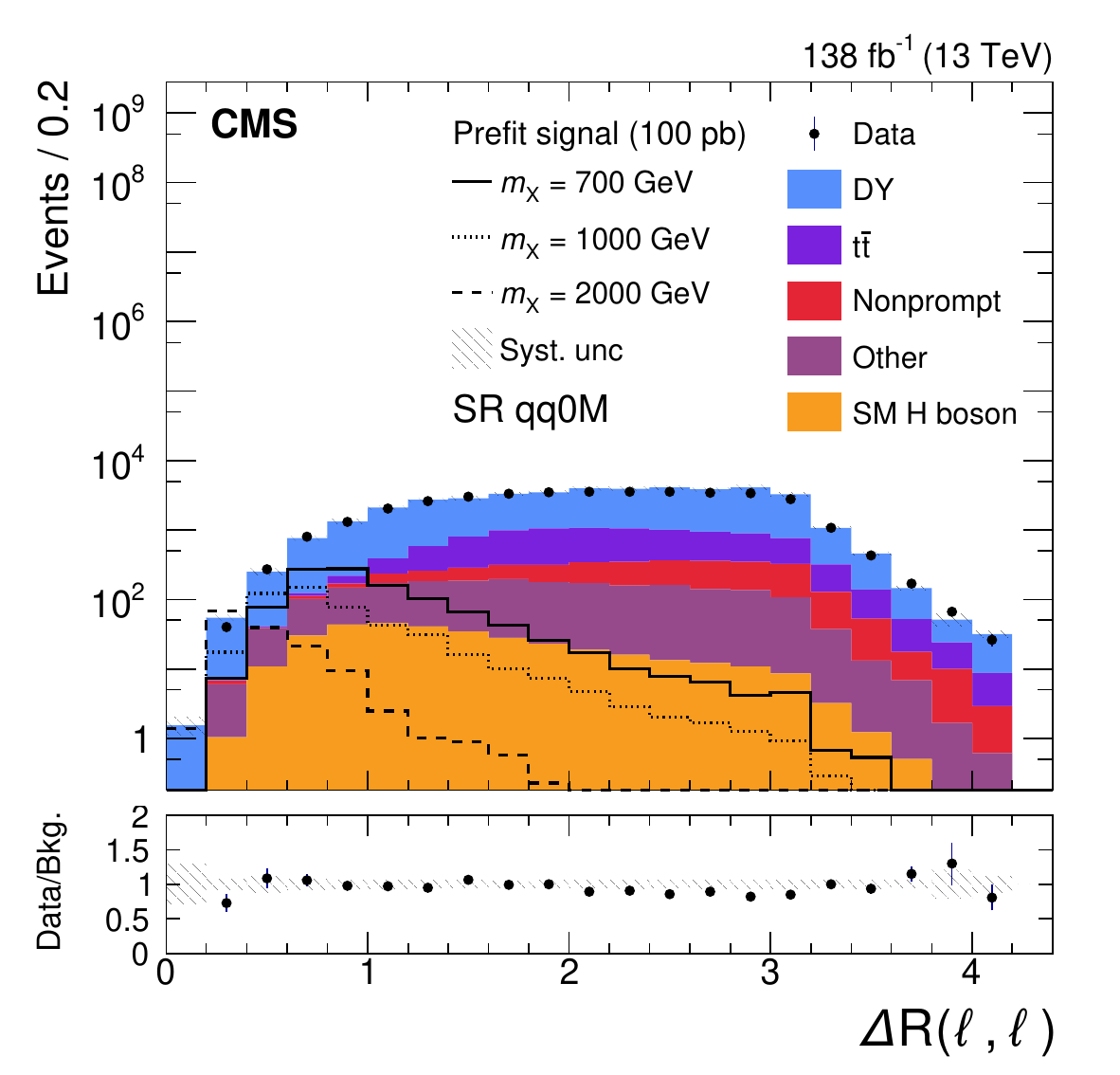}
  \includegraphics[width=0.48\textwidth]{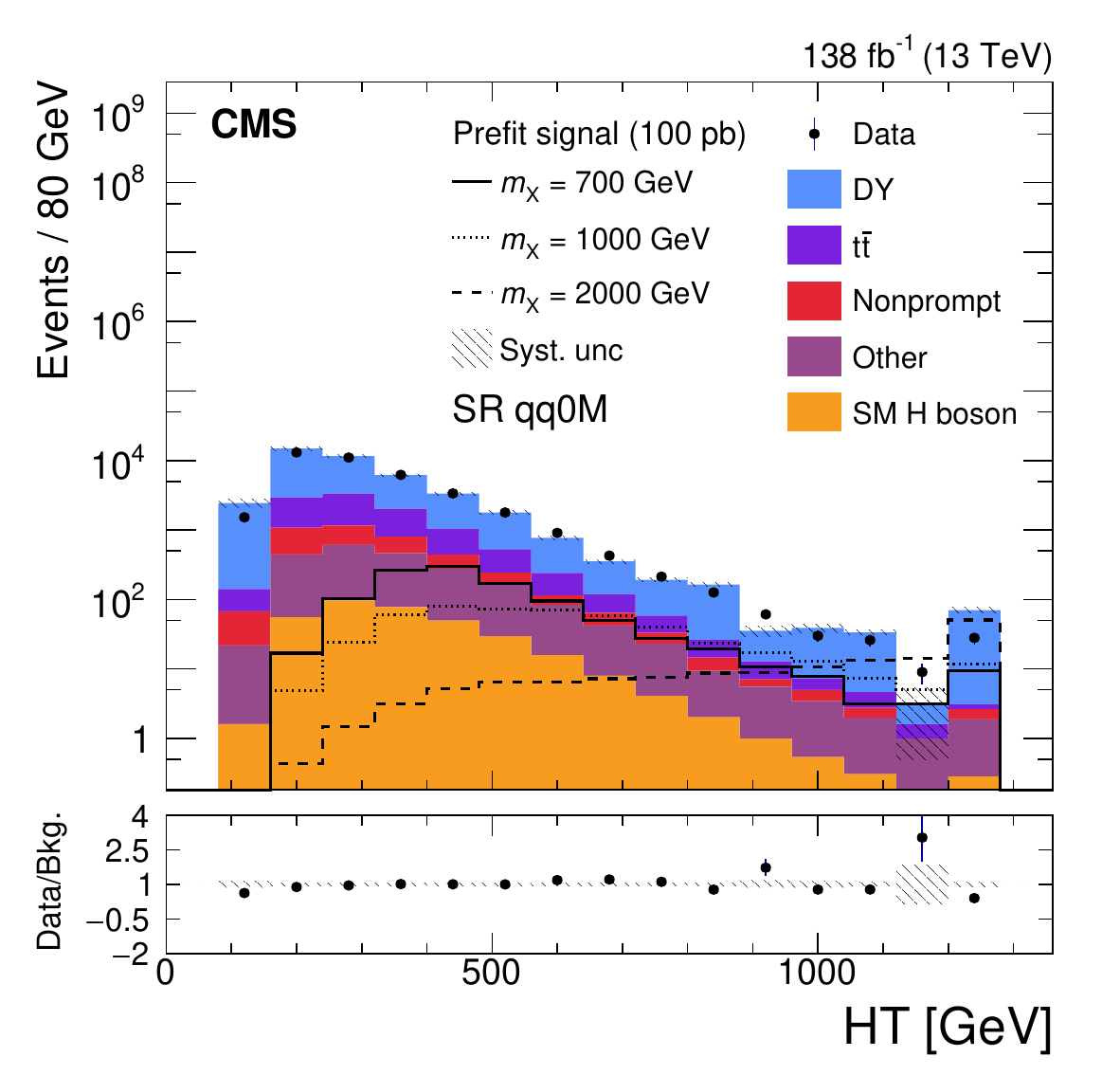}\\
  \includegraphics[width=0.48\textwidth]{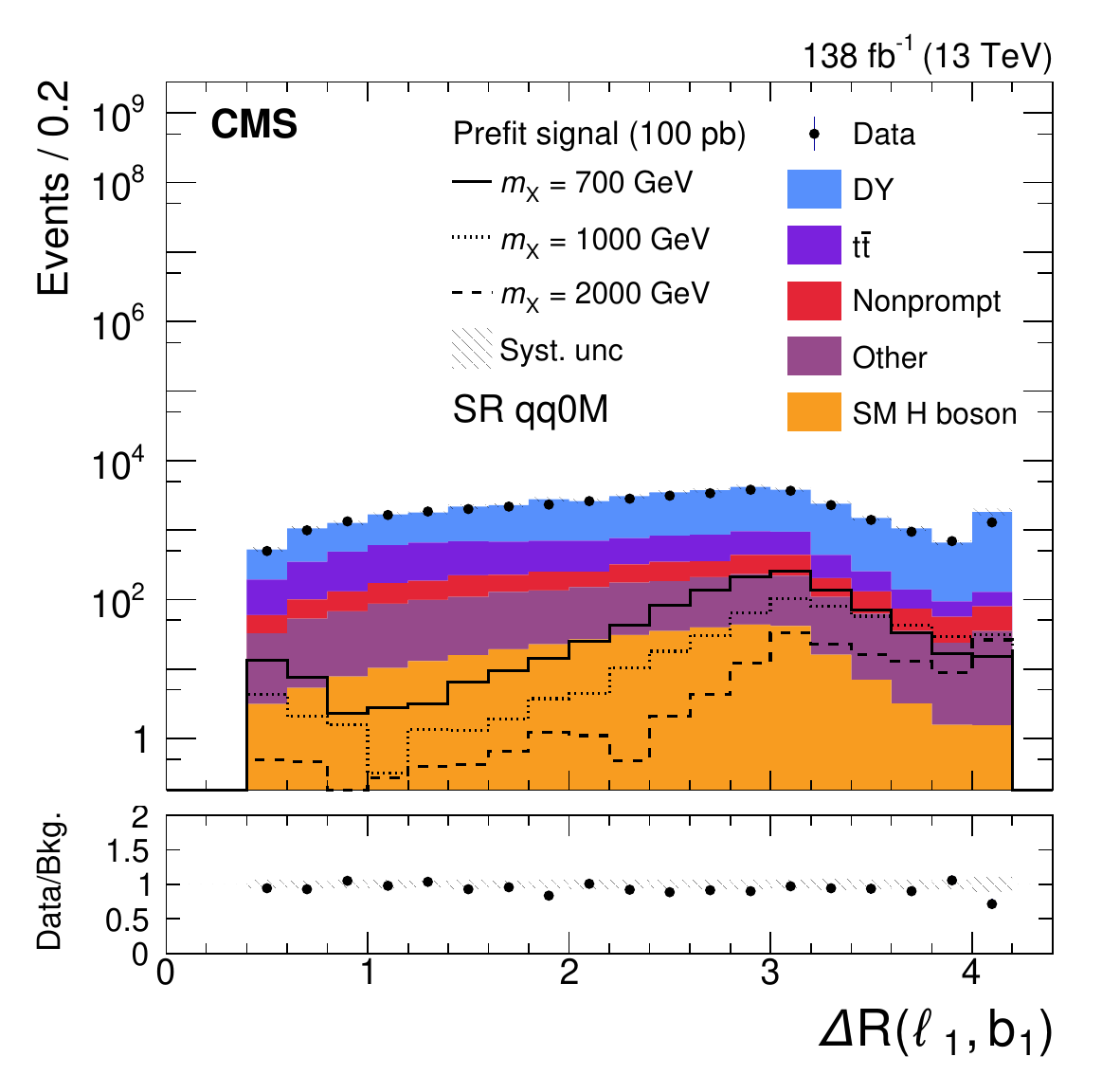}
  \includegraphics[width=0.48\textwidth]{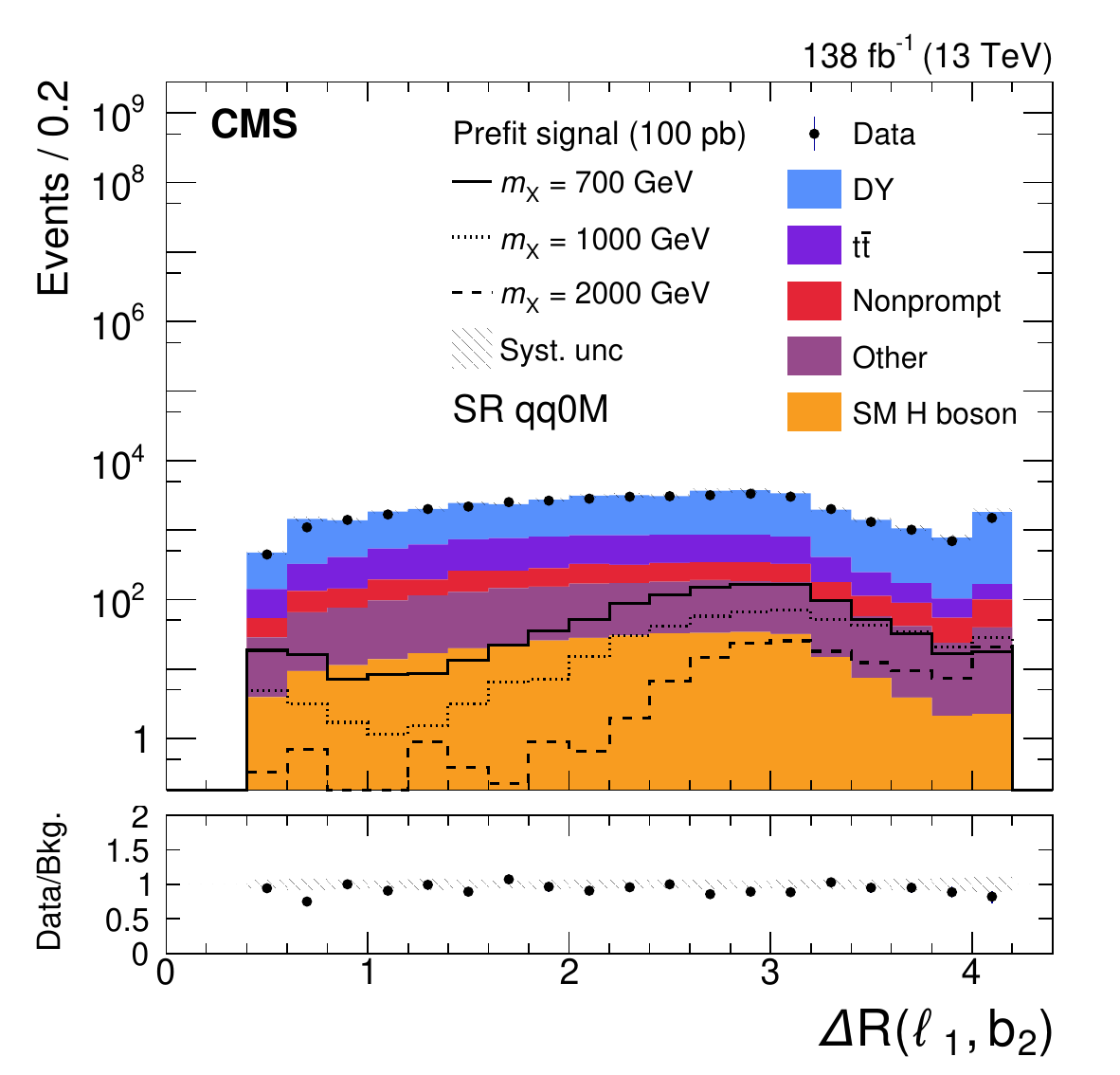}
  \caption{Prefit distributions of $\Delta R(\Pell,\Pell)$, HT, $\Delta R({\Pell}_1,{\PQb}_1)$, and $\Delta R({\Pell}_1,{\PQb}_2)$ (from upper left to lower right) in the SR qq0M, combining the electron and muon channels of \HY search and using all three data-taking years. Three histograms corresponding to resonance masses of $\mXres=700$, 1000, and 2000\GeV (all with $\mYres=125\GeV$) are also included in the plots, with the cross section of production of resonance \PX set to 100\unit{pb} and its branching fraction to \HY set to 1. The lower panel in each plot displays the ratio of the data to the total SM prediction, with the hatched bands representing the total systematic uncertainties in the backgrounds. The last bin includes the overflow.}
\label{fig:preBDT}
\end{figure*}

The kinematic variables of the final-state objects are used as the inputs of the BDT. After excluding correlated variables with correlation coefficients larger than 50\% in both signal and backgrounds samples (the correlation coefficients are determined using the TMVA package~\cite{Hocker:2007ht}), we use the following variables for training for the \HY signal: 
the leading lepton \pt and $\eta$, the dilepton \pt ($\pt^{\Pell\Pell}$), the \pt and $\phi$ of the leading jet from \PZ boson decay, the \pt of the two \PQb jets,
the \pt and invariant mass of the pair of \PQb-tagged jets, the scalar sum of the jet transverse momenta (HT), the spatial separation between two leptons [$\Delta R(\Pell,\Pell)$], the spatial separation between leading lepton and two \PQb jets [$\Delta R({\Pell}_1,{\PQb}_1)$, $\Delta R({\Pell}_1,{\PQb}_2)$], and the assigned resonance mass.
The variables used to train a BDT for the {\HH} signal are similar:
the leading lepton \pt and $\eta$, the $\pt^{\Pell\Pell}$, the \pt and $\phi$ of the leading jet from \PZ boson decay, the \pt of subleading jet from Higgs boson decay, 
the \pt and invariant mass of the pair of \PQb-tagged jets, the HT, the $\Delta R(\Pell,\Pell)$, the $\Delta R({\Pell}_1,{\PQb}_1)$, the $\Delta R({\Pell}_1,{\PQb}_2)$, 
and the difference between the reconstructed resonance mass and the assigned resonance mass. 
The variables $\Delta R(\Pell,\Pell)$, HT, $\Delta R({\Pell}_1,{\PQb}_1)$, and $\Delta R({\Pell}_1,{\PQb}_2)$ exhibit the strongest discriminating power among all those considered. The corresponding distributions are shown in Fig.~\ref{fig:preBDT}, where the electron and muon channels are combined and all three data-taking years are used.
The distributions of the SM backgrounds and BSM \HY signals use the default nuisance parameters (referred to as ``prefit'').
Three representative signal hypotheses are displayed, corresponding to resonance masses of $\mXres=700$, 1000, and 2000\GeV, respectively, with the mass of the \PY boson fixed at $\mYres=125\GeV$.
For visual clarity, these signal templates are normalized to a cross section of 100\unit{pb}, which allows for a direct comparison of the signal shapes relative to the background across the mass spectrum.

During the evaluation step, in contrast to the nominal parametrized BDT---where the hypothesized resonance mass is assigned to all MC (signal and background) and data events at each mass point---we instead assign all signal masses used in the BDT training to the background and data events (following the same method as that used during the BDT training). We refer to this procedure as the modified parametrized BDT.

In the nominal parametrized approach, the background BDT score (\bdtscore) varies with different resonance masses, requiring a large number of background histograms. In the modified parametrized approach, however, the background \bdtscore remains identical for a given BDT, which avoids generating an excessive number of background templates.

A comparison of the results obtained using these two evaluation methods shows that they are very similar, with the differences becoming negligible as the resonance mass increases.

\begin{figure*}[hbtp]\centering
  \includegraphics[width=0.48\textwidth]{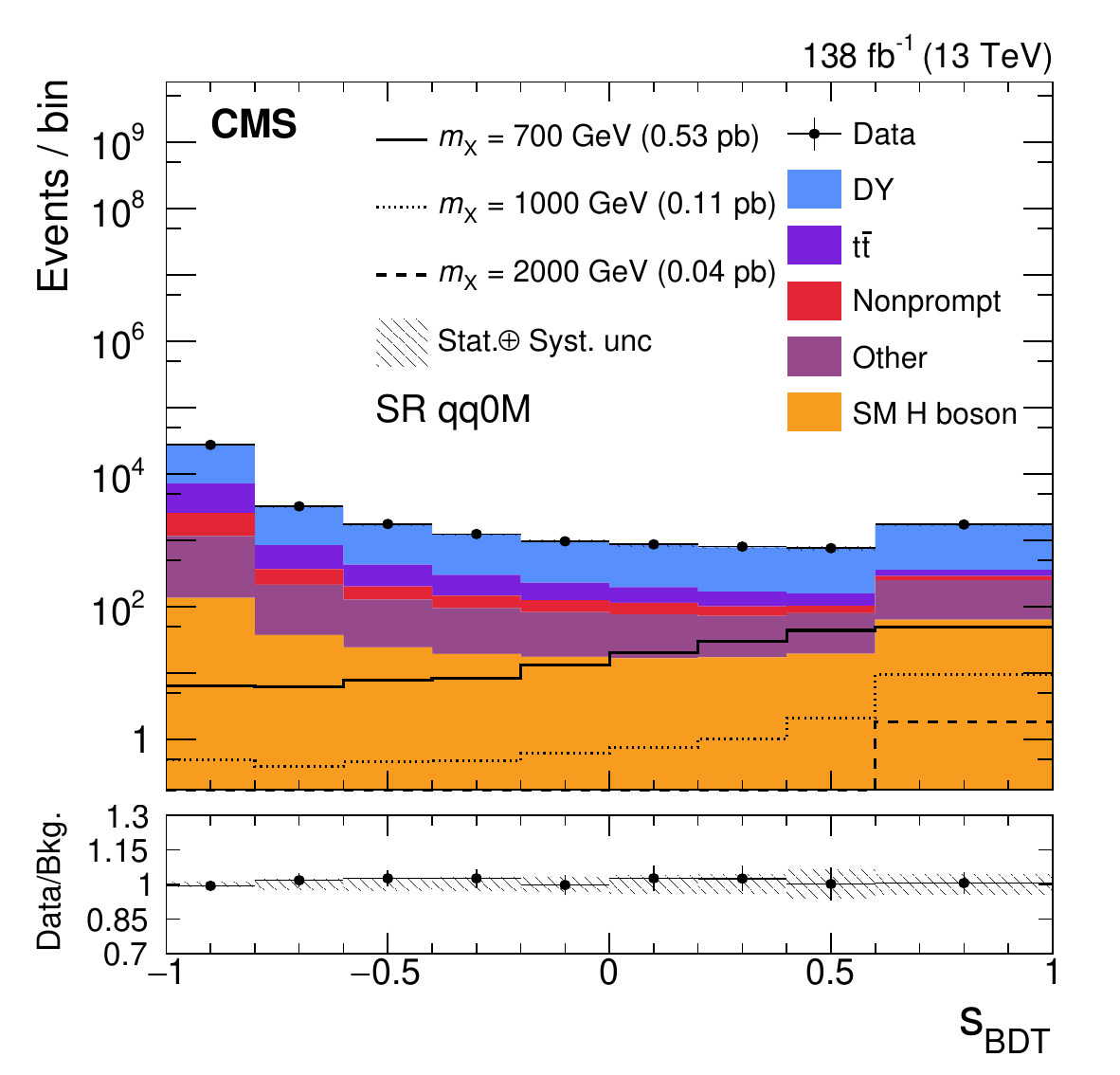}
  \includegraphics[width=0.48\textwidth]{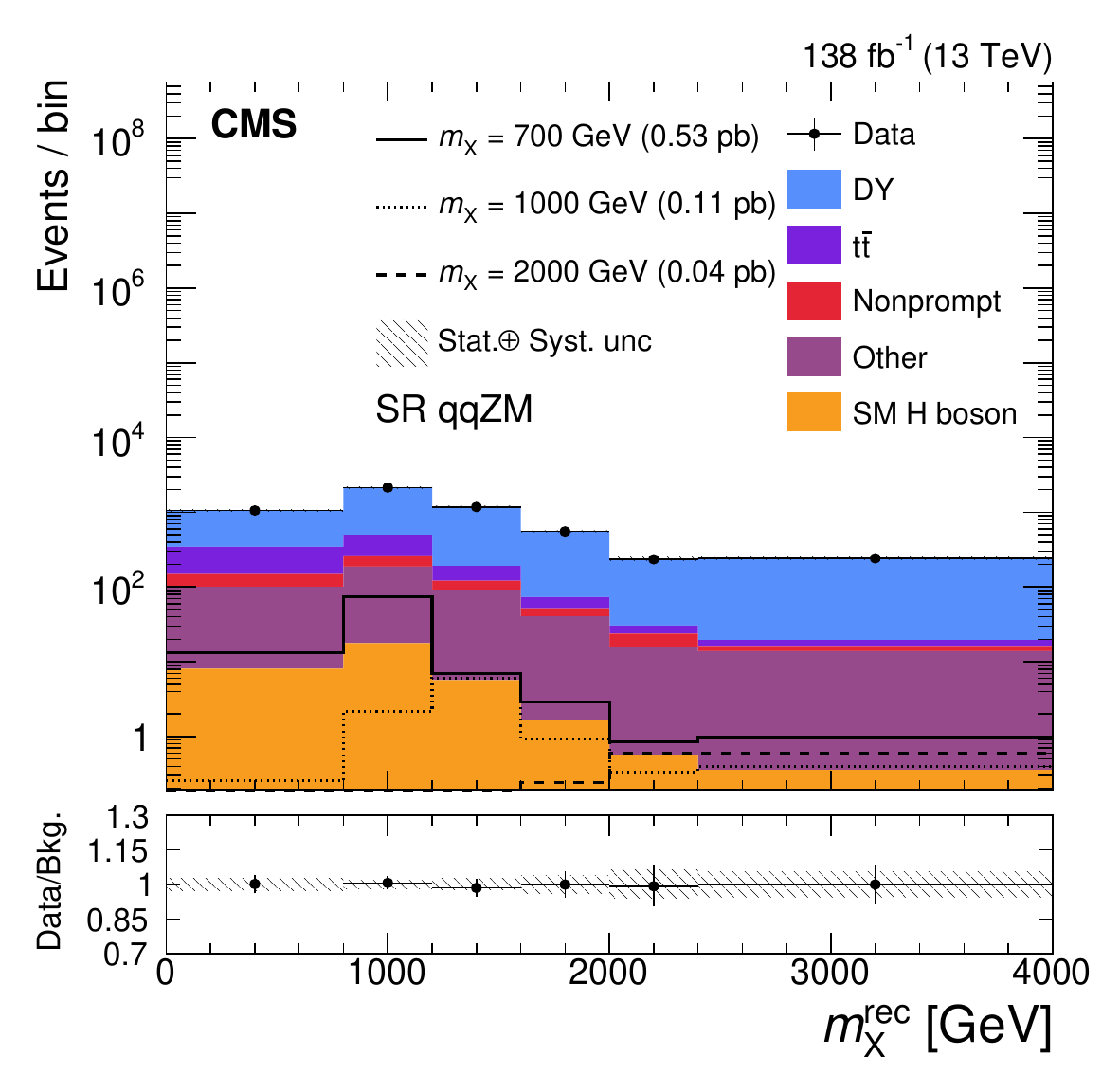}\\
  \includegraphics[width=0.48\textwidth]{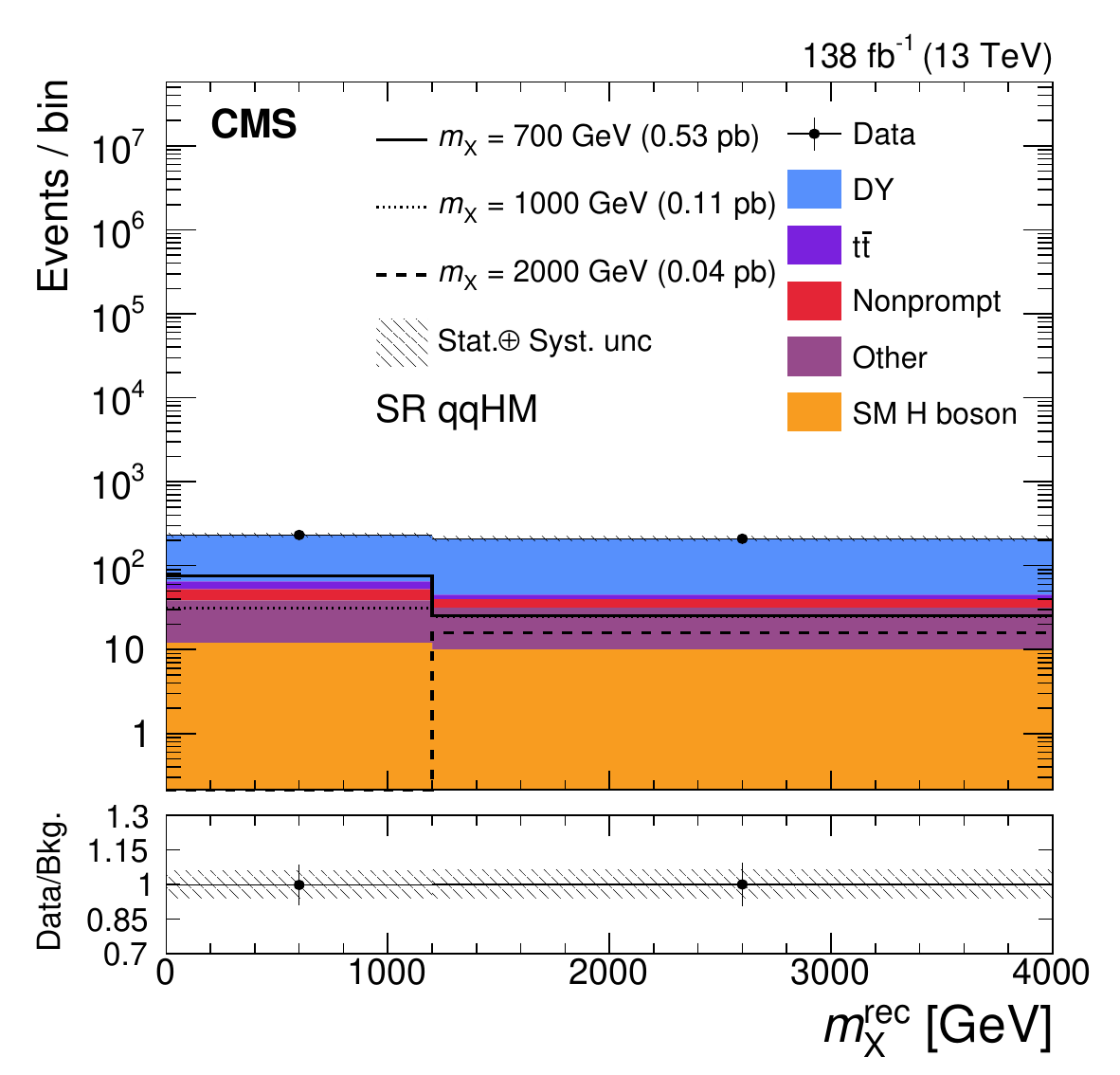}
  \includegraphics[width=0.48\textwidth]{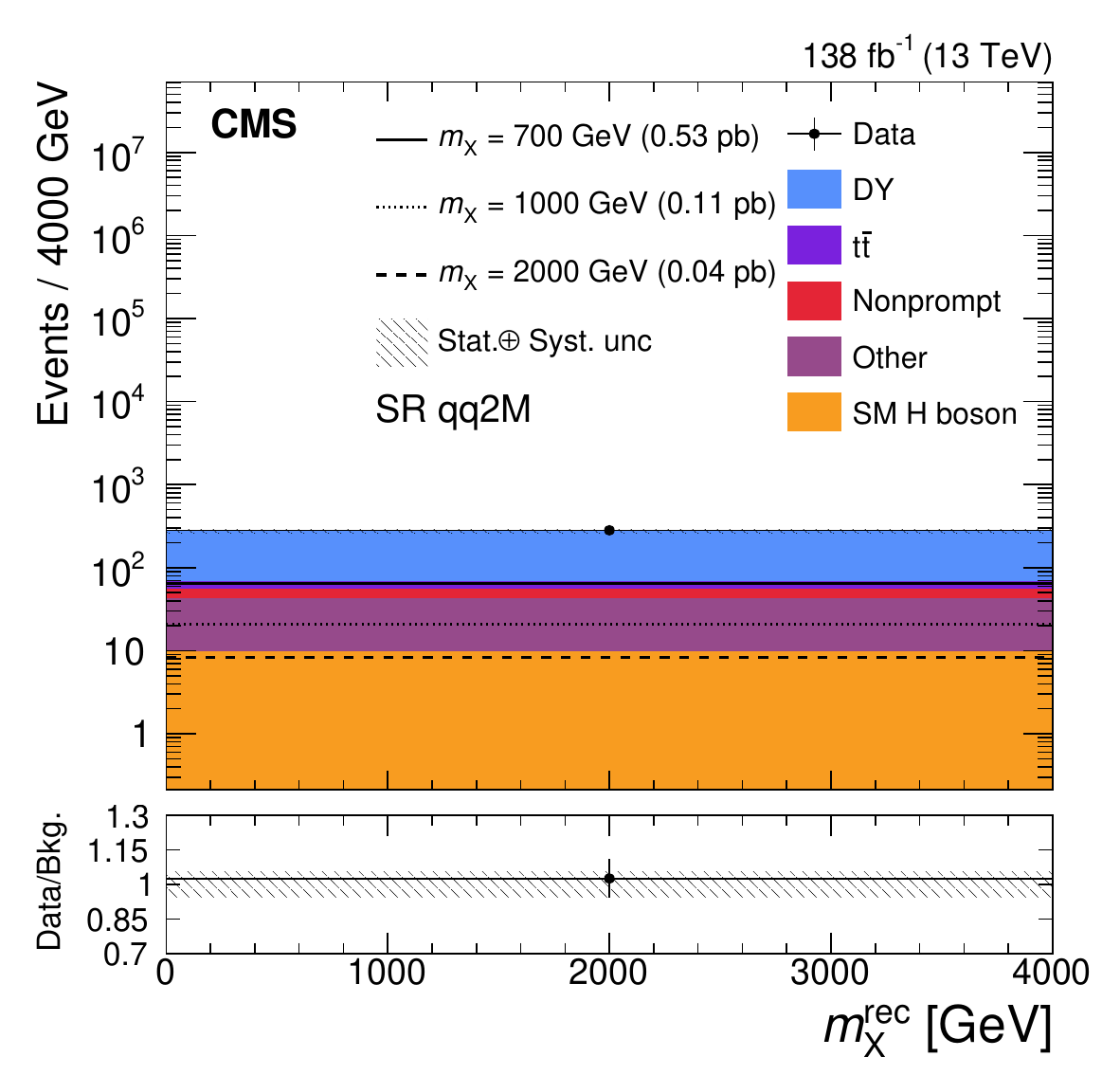}
  \caption{The distributions used for the fit in the four SRs of the \bbZZqq channel, combining the electron and muon channels of the \HY search are shown. 
The upper left plot displays the \bdtscore distribution in the SR qq0M.
The remaining three plots show the reconstructed resonance mass \mX distributions in the qqZM, qqHM, and qq2M SRs, from upper right to lower right.
Three histograms corresponding to resonance masses of $\mXres=700$, 1000, and 2000\GeV are also included in the plots, with the \ppHY production cross section set to its respective observed upper limit. These histograms account for the branching fractions of the \zzqqll decays. 
The lower panel in each plot displays the ratio of the data to the total SM prediction, with the hatched bands representing the overall uncertainty in the combined background expectations. The histograms of backgrounds are the postfit ones.}
\label{fig:qq_SR}
\end{figure*}

\begin{figure}[hbtp]\centering
  \includegraphics[width=0.48\textwidth]{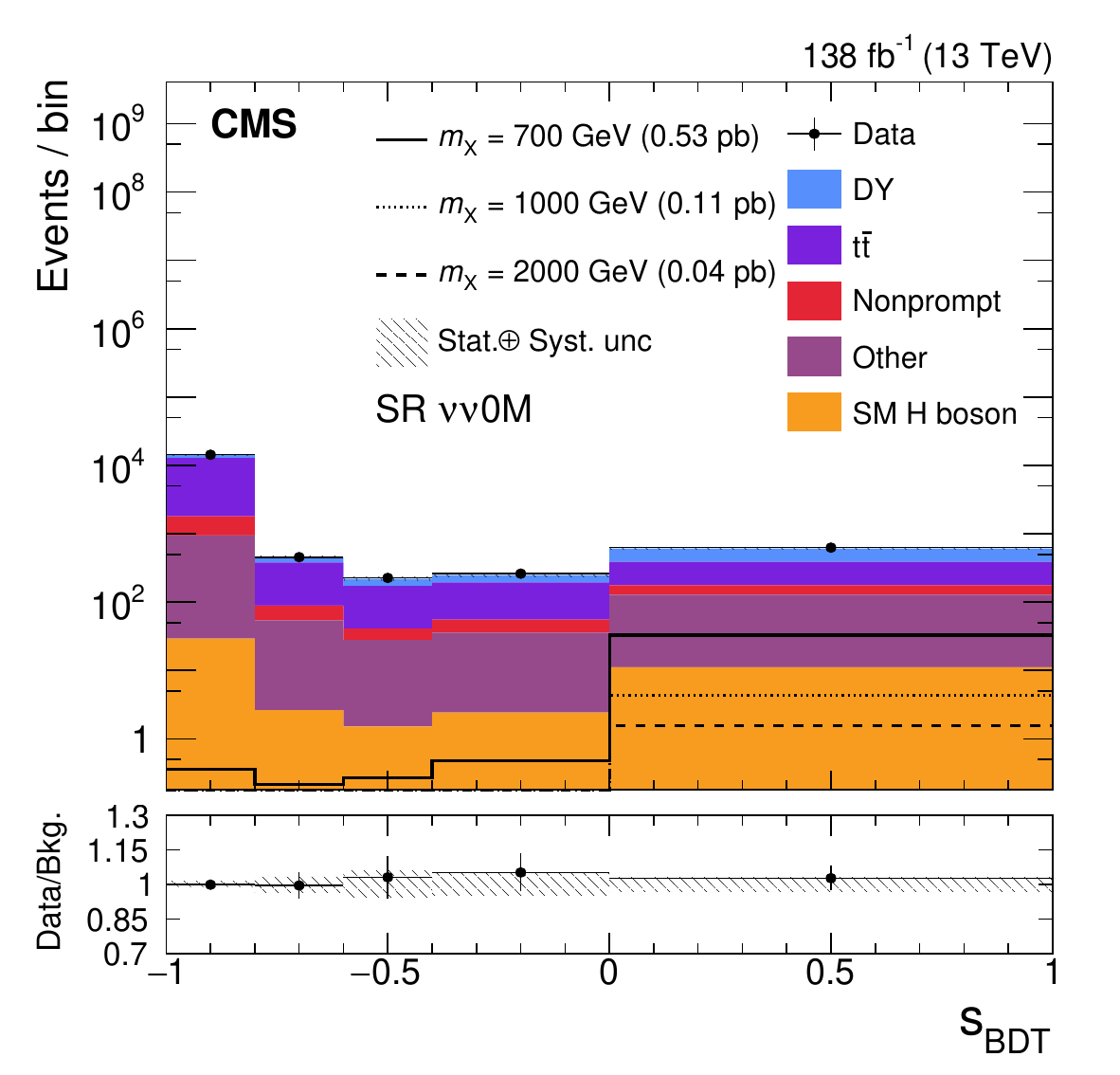}
  \includegraphics[width=0.48\textwidth]{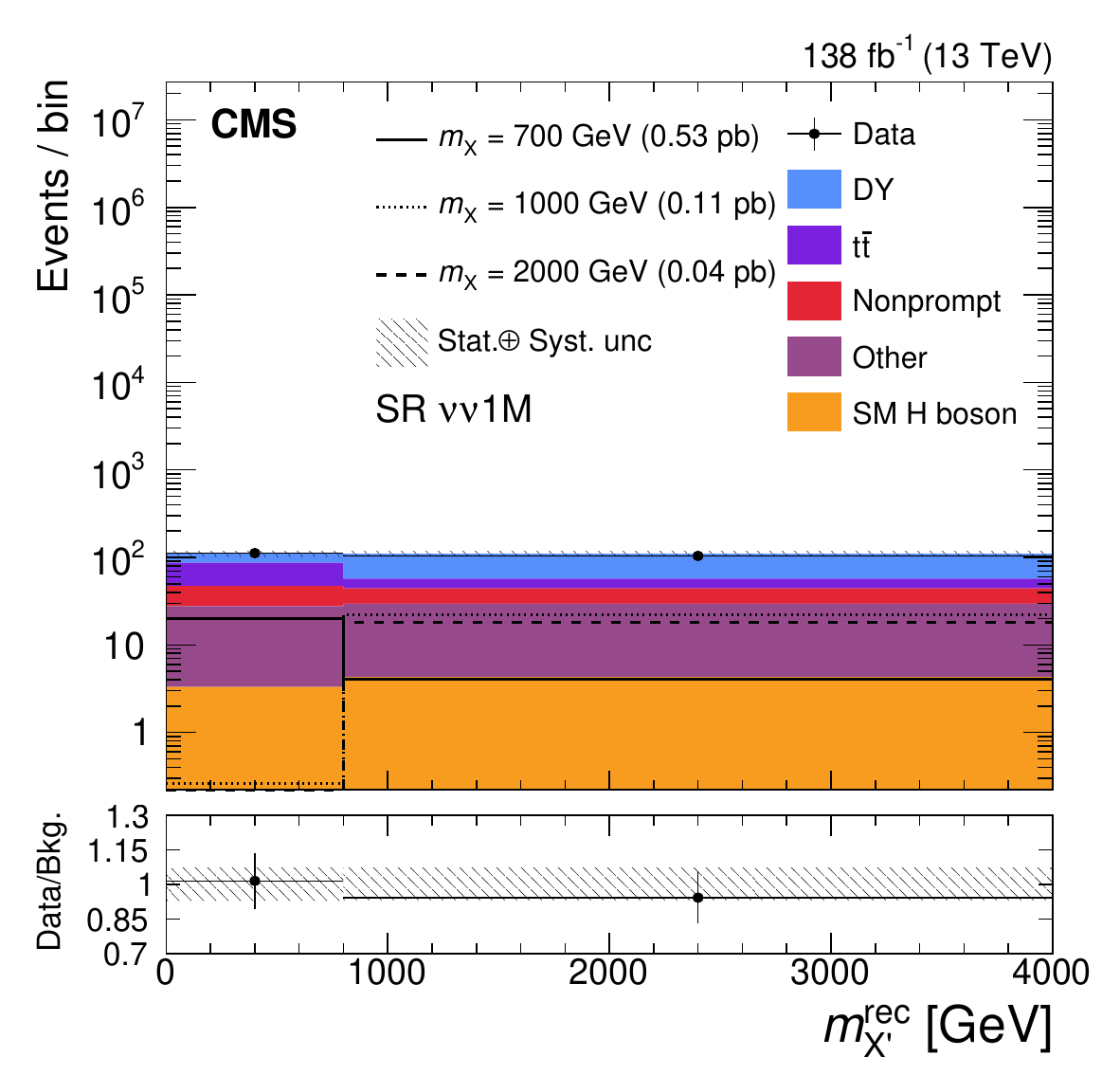}
  \caption{The distributions used for the fit in the two SRs of the \bbZZvv channel, combining the electron and muon channels of the \HY search are shown.
\cmsLeft: the \bdtscore distribution in the SR $\PGn\PGn$0M. \cmsRight: the reconstructed resonance mass \mXp distributions in the SR $\PGn\PGn$1M.
Three histograms corresponding to resonance masses of $\mXres=700$, 1000, and 2000\GeV are also included in the plots, with the \ppHY production cross section set to its respective observed upper limit. These histograms account for the branching fractions of the \zzllvv decays. 
The lower part of each plot displays the ratio of the data to the total SM prediction, with the hatched bands representing the overall uncertainty in the combined background expectations. The histograms of backgrounds are the postfit ones.}
\label{fig:vv_SR}
\end{figure}

The comparisons between data and SM predictions in the four SRs of the \bbZZqq channel of \HY signals are shown in Fig.~\ref{fig:qq_SR}. 
Figure~\ref{fig:vv_SR} shows the corresponding comparison in the two SRs of the \bbZZvv channel.
Three representative signal distributions, corresponding to resonance masses of $\mXres=700$, 1000, and 2000\GeV (all with $\mYres=125\GeV$), are also displayed. For each mass hypothesis, the \ppHY production cross section is normalized to its respective observed upper limit.
The SM branching fractions for \zzqqllvv are applied for sequential decays of the \PZ bosons.

For the \HH signals, the analysis procedure is nearly identical, with the only difference being the use of a different BDT in the fully resolved SRs.
In addition, the branching fractions for all SM particles, namely the \PH and \PZ bosons, are set to their respective SM values.

Because of the negative weights in the NLO MC samples---which arise from the cancellation of infrared divergences in the matrix element calculation---and the limited number of MC events, the histogram binning is chosen such that all SM processes, particularly the dominant DY background, yield positive event counts in each bin.
The histograms of the SM processes in these figures are normalized according to a fit to data (referred to as ``postfit'') under the hypothesis that there is no BSM signal.

\section{Systematic uncertainties}
Systematic uncertainties affecting the measurements arise from both experimental and theoretical sources.
Experimental uncertainties stem from detector effects and the methods used to compute higher-level quantities, such as reconstruction efficiencies. 
Theoretical uncertainties originate from variations in inputs, including the choice of renormalization (\muR) and factorization (\muF) scales for the hard process and the parton showering, as well as the choice of PDFs.
Each systematic uncertainty is quantified by evaluating its effect on the yield and shape of kinematic distributions in the signal and background processes. 

\subsection{Experimental uncertainties}
The uncertainties associated with lepton reconstruction and selection efficiencies are measured to be at the percent level. 
The uncertainty in the trigger efficiencies is 1\%--8\% depending on the lepton $\eta$ and the $\Delta R$ of two leptons. These uncertainties are treated as correlated across the three years of data taking.
During the 2016--2017 data taking, a gradual shift in the timing of the inputs of the ECAL hardware level trigger in the region of $\abs{\eta} > 2$ caused a specific trigger inefficiency. Event-level weight corrections and their uncertainties are included in the simulation to account for this. 
The uncertainties in the jet energy scale (JES) and jet energy resolution (JER) are evaluated using simulated events by rescaling and spreading the jet \pt. 
These uncertainties are propagated to the final results by introducing the corresponding nuisance histograms of the fit variables, with the varied JES or JER corrections~\cite{jer} applied.
We consider different sources of JES uncertainty; some are fully or partially correlated across different data-taking periods, while others are treated as uncorrelated. 
The uncertainties in JES and JER, as well as the variation of the unclustered energy, are propagated to \ptmiss.

Systematic uncertainties arise from the selection of \PQb jets and the corresponding efficiencies. The \PQb jet selection uncertainties are split into different sources, 
depending on the jets they affect: jets originating from \PQb quarks, \PQc quarks, or light-flavor quarks or gluons. 
Furthermore, these are split into the statistical and systematic components, where the former are not correlated between eras of data taking, but the latter are. 
Systematic uncertainties in the \textsc{PNet} tagging of AK8 jets arise from various sources, which affect the measured tagging efficiency. 
Similarly, they can be categorized into systematic and statistical components and follow the same correlation strategy as \PQb tagging of AK4 jets.

The integrated luminosities for the 2016, 2017, and 2018 data-taking years have 1.2--2.5\% individual uncertainties~\cite{CMS:2021xjt,CMS-PAS-LUM-17-004,CMS-PAS-LUM-18-002}, while the overall uncertainty for the 2016--2018 period is 1.6\%. 
The total inelastic cross section of 69.2\unit{mb}, used to correct the pileup distribution in simulation, is varied by $\pm$4.6\%~\cite{pileup}. This uncertainty is treated as correlated across the different data-taking years.

The uncertainty in the nonprompt-lepton background estimation has both statistical and systematic components.
The statistical uncertainty originates from the limited yields in the sideband regions and is propagated through the ratio ($\Ni/\Nii$).
The systematic uncertainty is defined as the difference between the nonprompt-lepton background yields obtained from the method based on control samples in data and those estimated using simulated samples, as described earlier.

We derive the uncertainties in the normalization factors for the DY and \ttbar processes from the fits described earlier. The corresponding uncertainties amount to a few percent of the normalization values. 
For other backgrounds estimated using simulation, the normalization uncertainties are based on the cross sections, which are taken either from CMS measurements or from higher-order theoretical calculations.
The cross section uncertainties for diboson and triboson processes range from a few percent to about 10\%, while those for less significant processes range from a few percent up to several tens of percent.

\subsection{Theoretical uncertainties}
We estimate the theoretical uncertainties for simulated processes by varying the \muR and \muF scales up and down independently by a factor of 2 relative to their nominal value in each event, subject to the condition $0.5 \leq \mu_\mathrm{R}/\mu_\mathrm{F} \leq 2$~\cite{Cacciari:2004,Catani:2003}.
The scale used in the parton showering in the signal samples has negligible impact, whereas a 2\% uncertainty is assigned to the background processes.
The uncertainty is derived from the resulting variations in the $\pt^{\Pell\Pell}$ distribution, with the largest deviation from the nominal shape taken as the uncertainty.
We estimate the uncertainties in the PDFs by combining the contributions from replicas of the NNPDF3.1 PDF set, following the procedure outlined in Ref.~\cite{Butterworth_2016}. 
These theoretical uncertainties are applied to the DY and \ttbar backgrounds, as well as to the signal process. For the DY and \ttbar backgrounds, we take into account both normalization and shape differences, whereas for the signal process, we consider only shape differences.

The uncertainties in the integrated luminosity, nonprompt-lepton background estimation, the normalization factor of the \ttbar contribution, and the cross sections of the simulations (excluding DY and \ttbar) are treated as affecting only the normalization in the fit. 
We use log-normal probability density functions to model these nuisance parameters.
We treat the theoretical uncertainties in the signal process as shape-only uncertainties. For each uncertainty, three one-dimensional binned histograms are provided: one representing the nominal kinematic distributions and two corresponding to the $\pm 1\sigma$ variations. We then estimate the shape uncertainty through vertical interpolation among these three histograms.
We treat all other uncertainties as affecting both the shape and normalization during the fit. The shape component is modeled using vertical interpolation, while the normalization component is described by log-normal probability density functions.

\section{Results}

We perform a simultaneous binned profile likelihood fit across all SRs, comparing the signal and background templates to the observed data.
In the qq0M and $\PGn\PGn$0M SRs, the fit uses the \bdtscore as the discriminating variable, while in the remaining SRs, it uses resonance mass \mX or \mXp.

The cross sections for the \ppHYbbZZ and \ppHH production processes are each set to a fixed value for all resonance mass hypotheses.
For both the \ppHY and \ppHH signals, the branching fractions of the SM Higgs boson with a mass of 125\GeV are used for the \PH boson decay.
For the \ppHY signal, the cross section of \bbWW is obtained by multiplying the cross section with the ratio of the branching fractions $\mathcal{B}(\PY\to \PW\PW)$ to $\mathcal{B}(\PY\to \PZ\PZ)$~\cite{Heinemeyer:1559921}.
The subsequent decay branching fractions of the vector bosons are also taken into account.

The data are consistent with the SM prediction, as shown in Figs.~\ref{fig:qq_SR} and~\ref{fig:vv_SR}.
Upper limits on the production cross sections of the \ppHYbbZZ and \ppHH processes as functions of the resonance mass are set using the CMS statistical analysis tool \combine~\cite{combine}, which is based on the \textsc{RooFit}~\cite{Verkerke:2003ir} and \textsc{RooStats}~\cite{Moneta:2010pm} frameworks. 
The upper limits are calculated at 95\% confidence level (\CL) using a modified frequentist method~\cite{CLS2,CLS1} within an asymptotic approximation~\cite{Cowan:2010js}.

\begin{figure}[htbp]\centering
  \includegraphics[width=0.48\textwidth]{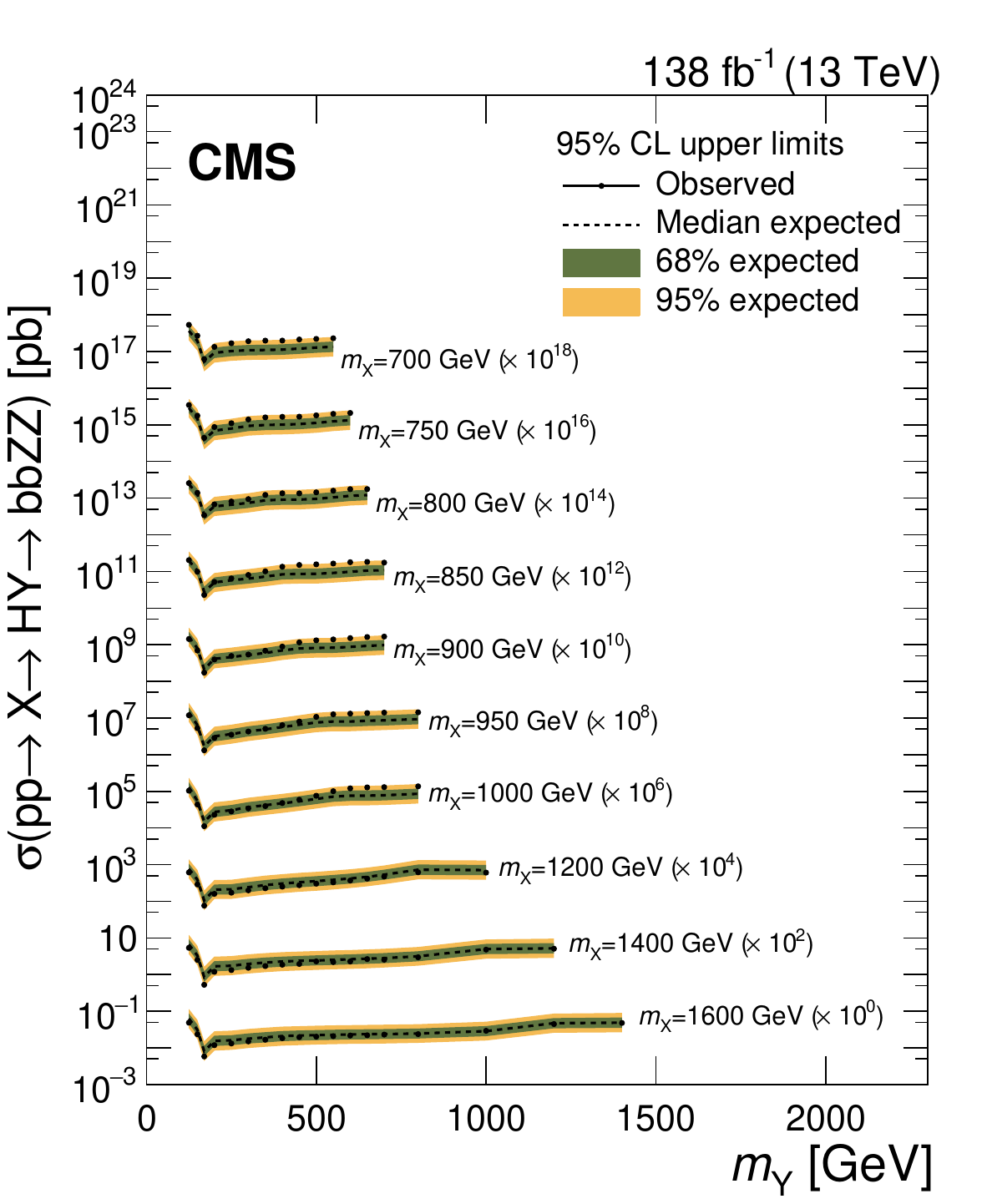}
  \includegraphics[width=0.48\textwidth]{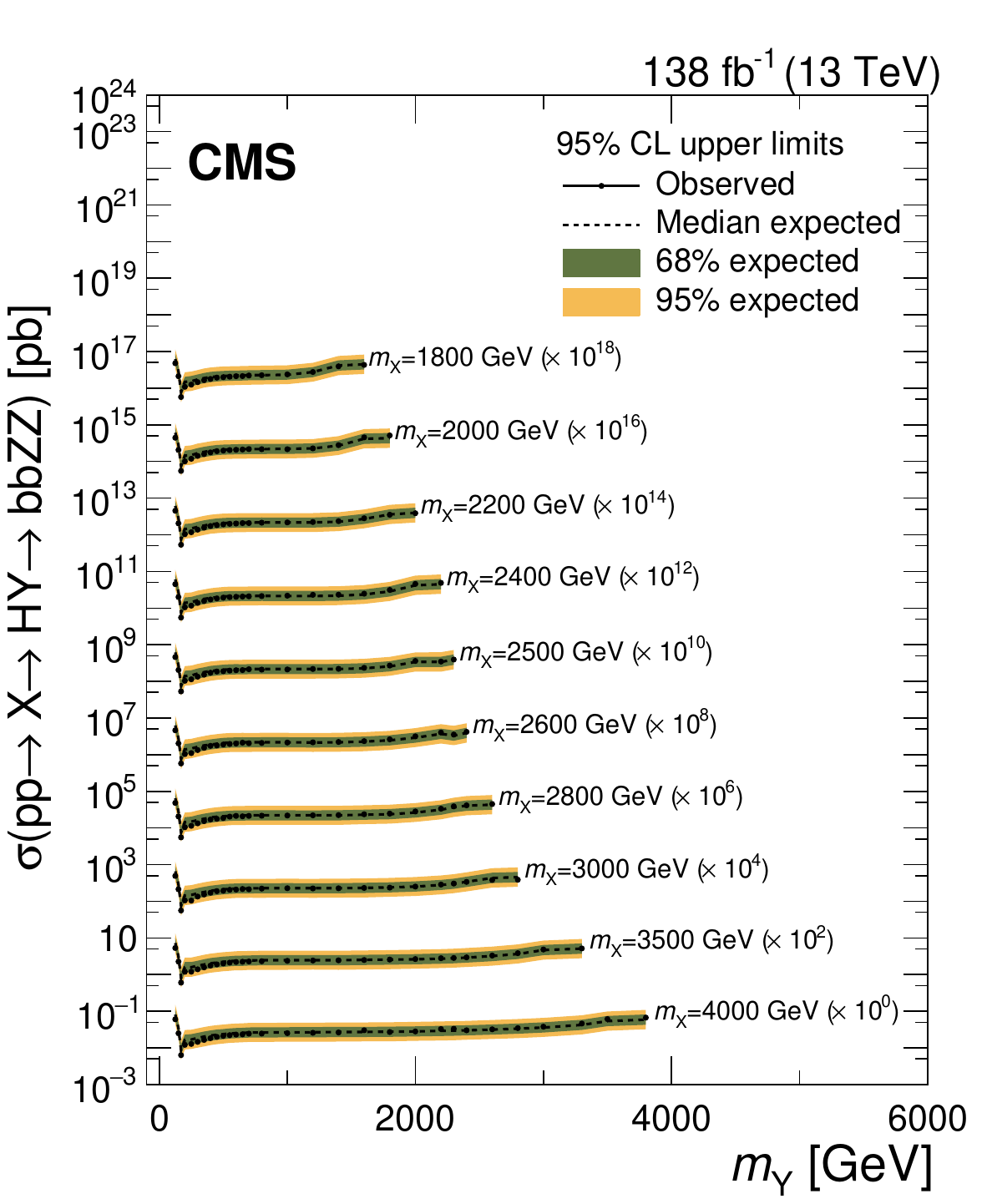}
  \caption{Upper limits on the production cross section of \ppHYbbZZ in the two-dimensional parameter space of the masses of the two BSM scalars ($m_\PX$, $m_\PY$) combining all the SRs as well as the electron and muon channels. The inner (green) band and the outer (yellow) band indicate the regions containing 68\% and 95\%, respectively, of the distribution of limits expected under the background-only hypothesis.}
\label{fig:limitHY}
\end{figure}

\begin{figure}[htb]\centering
  \includegraphics[width=\cmsFigWidth]{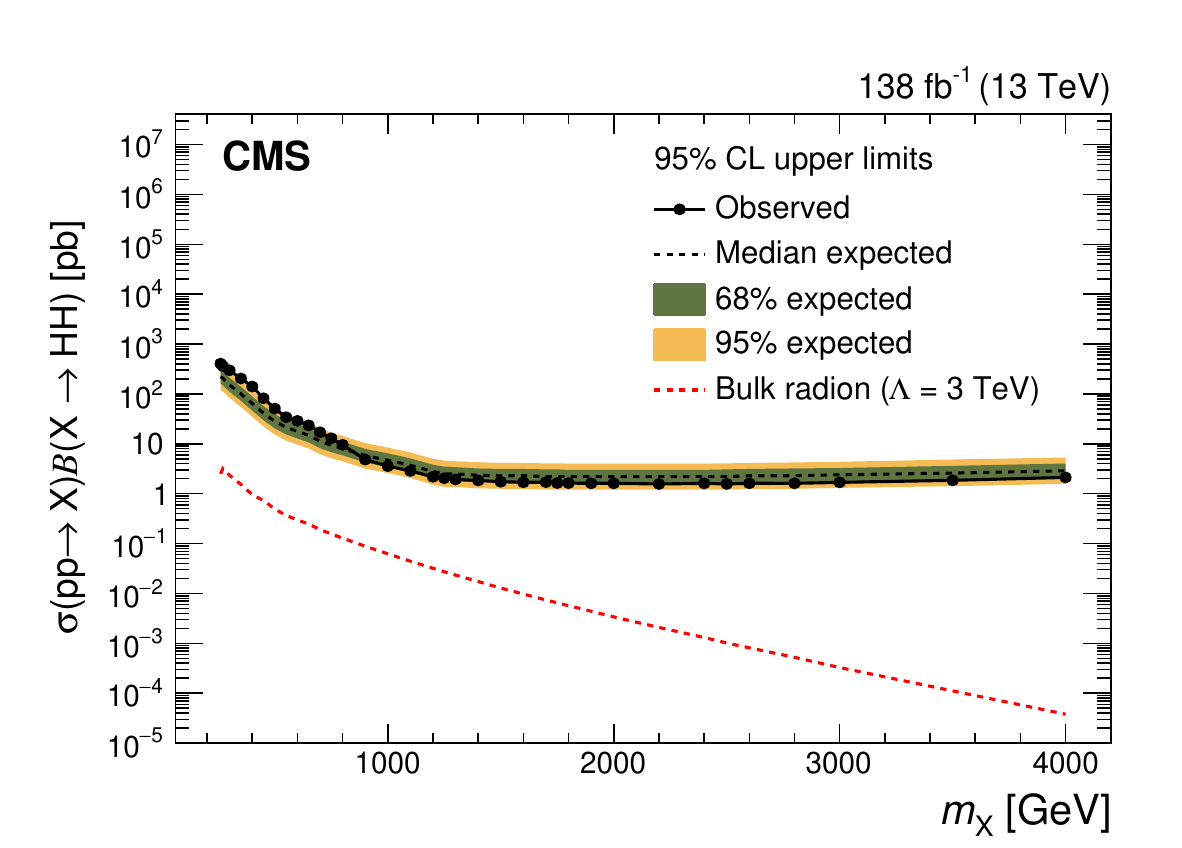}
  \caption{Upper limits on the production cross section of \ppHH with respect to the resonance mass $m_{\PX}$, combining all the SRs as well as the electron and muon channels, and taking into account the theoretical prediction of the branching fraction of the resonance to \HH. The inner (green) band and the outer (yellow) band indicate the regions containing 68\% and 95\%, respectively, of the distribution of limits expected under the background-only hypothesis. The theoretical values (red dashed line) are also provided on the plot.}
\label{fig:limit}
\end{figure}

The systematic uncertainties are modeled as nuisance parameters in the maximum likelihood fits to determine the best-fit signal strength. These nuisance parameters are allowed to modify both the shape and normalization of the backgrounds and signal in the fit. 

Figure~\ref{fig:limitHY} shows the upper limits on the cross section for \ppHYbbZZ production in the two-dimensional phase space of the masses of the two BSM scalars ($m_\PX$, $m_\PY$), with upper limits varying from 5 to 500\unit{fb}, depending on the scalar mass hypotheses. 
The results are comparable to the sensitivity achieved in other analyses, which focus on \PH decays to $\gamma\gamma$ or $\Pgt\Pgt$ and \PY decays into a pair of bottom quarks or massive vector bosons~\cite{2025368,2023atlasvvtata,atlasbbgg}. 
As a by-product of this search, the results are also interpreted in the context of resonant \HH production. Figure~\ref{fig:limit} shows the upper limits on the product of resonance production cross section and the branching fraction to \HH for values of the resonance mass from 260\GeV to 4\TeV, from the combination of all the SRs, together with both the electron and muon channels. The upper limits decrease from 400 to 1\unit{pb} as the resonance mass increases.
The theoretical value of the product of the cross section of the bulk radion production and its branching fraction to \HH is also provided~\cite{xandatheo}.
The similar study~\cite{PhysRevD.102.032003} searching for resonant \HH  conducted by CMS using the 2016 dataset limits on its production up to a mass of 1\TeV.  This earlier study achieved limits several times more stringent that those presented here in the low-mass resonance region, but becomes comparable at 1\TeV.
A key distinction lies in the treatment of the low-\mll regime; while the previous study employed a much looser requirement of $m_{\Pell\Pell} > 15\GeV$ in the \bbZZqq channel, the present analysis adopts a more restrictive threshold to ensure orthogonality to the \bbWW analysis. In this low-\mll kinematic regime, the \bbWW contribution dominates over \bbZZf. 
Therefore, this cut is implemented to avoid signal overlap and to maintain a clear definition of the signal, leaving the lower-mass regime to be more appropriately addressed by a dedicated \bbWW study. Taking the resonance mass at 1\TeV as a representative case, this threshold retains approximately 50\% of the \bbZZf signal while effectively eliminating \bbWW contamination. 
Crucially, the selection suppresses the \ttbar background by nearly 70\%, ensuring a well-defined \bbZZf SR. Such orthogonality is essential for a clear physical interpretation of the results and is required to facilitate future statistically independent combinations between the \bbZZf and \bbWW channels, especially in the \HY study.
Tabulated results are provided in the HEPData record for this analysis~\cite{hepdata}.

\section{Summary}
A search for a new resonance \PX decaying into either a Higgs boson and a new scalar boson \PY (\HY) or a pair of Higgs bosons (\HH) has been conducted in this study, using proton-proton collision data collected by the CMS experiment from 2016 to 2018 at $\sqrt{s}=13\TeV$, corresponding to an integrated luminosity of 138\fbinv.
The events of interest are characterized by two b quarks from the decay of one Higgs boson, as well as two leptons and either two additional quarks or two neutrinos from two \PZ bosons produced by the decay of the other \PY or the \PH.
No significant deviations are observed in the signal regions between the data and standard model predictions.
For resonant \HY production, upper limits are set on the cross section for the process \ppHYbbZZ in the two-dimensional parameter space of the masses \mXres and \mYres.
These limits vary from 500 to 5\unit{fb}, with the most stringent constraints achieved for high-mass \PX resonances and \PY masses near the threshold for decay into a pair of massive vector bosons.
These are comparable to the limits achieved in other analyses, which focus on \PH decays to $\PGg\PGg$ or $\Pgt\Pgt$ and \PY decays into a pair of bottom quarks or massive vector bosons.
For resonant \HH production, upper limits are placed on the cross section for the process \ppHH as a function of the resonance mass, varying from 400 to 1\unit{pb} as \mXres increases.

\begin{acknowledgments}
We congratulate our colleagues in the CERN accelerator departments for the excellent performance of the LHC and thank the technical and administrative staffs at CERN and at other CMS institutes for their contributions to the success of the CMS effort. In addition, we gratefully acknowledge the computing centers and personnel of the Worldwide LHC Computing Grid and other centers for delivering so effectively the computing infrastructure essential to our analyses. Finally, we acknowledge the enduring support for the construction and operation of the LHC, the CMS detector, and the supporting computing infrastructure provided by the following funding agencies: SC (Armenia), BMBWF and FWF (Austria); FNRS and FWO (Belgium); CNPq, CAPES, FAPERJ, FAPERGS, and FAPESP (Brazil); MES and BNSF (Bulgaria); CERN; CAS, MoST, and NSFC (China); MINCIENCIAS (Colombia); MSES and CSF (Croatia); RIF (Cyprus); SENESCYT (Ecuador); ERC PRG and PSG, TARISTU24-TK10 and MoER TK202 (Estonia); Academy of Finland, MEC, and HIP (Finland); CEA and CNRS/IN2P3 (France); SRNSF (Georgia); BMFTR, DFG, and HGF (Germany); GSRI (Greece); NKFIH (Hungary); DAE and DST (India); IPM (Iran); SFI (Ireland); INFN (Italy); MSIT and NRF (Republic of Korea); MES (Latvia); LMTLT (Lithuania); MOE and UM (Malaysia); BUAP, CINVESTAV, CONACYT, LNS, SEP, and UASLP-FAI (Mexico); MOS (Montenegro); MBIE (New Zealand); PAEC (Pakistan); MES, NSC, and NAWA (Poland); FCT (Portugal); MESTD (Serbia); MICIU/AEI and PCTI (Spain); MOSTR (Sri Lanka); Swiss Funding Agencies (Switzerland); MST (Taipei); MHESI (Thailand); TUBITAK and TENMAK (T\"{u}rkiye); NASU (Ukraine); STFC (United Kingdom); DOE and NSF (USA).

\hyphenation{Rachada-pisek} Individuals have received support from the Marie-Curie program and the European Research Council and Horizon 2020 Grant, contract Nos.\ 675440, 724704, 752730, 758316, 765710, 824093, 101115353, 101002207, 101001205, and COST Action CA16108 (European Union); the Leventis Foundation; the Alfred P.\ Sloan Foundation; the Alexander von Humboldt Foundation; the Science Committee, project no. 22rl-037 (Armenia); the Fonds pour la Formation \`a la Recherche dans l'Industrie et dans l'Agriculture (FRIA) and Fonds voor Wetenschappelijk Onderzoek contract No. 1228724N (Belgium); the Beijing Municipal Science \& Technology Commission, No. Z191100007219010, the Fundamental Research Funds for the Central Universities, the Ministry of Science and Technology of China under Grant No. 2023YFA1605804, the Natural Science Foundation of China under Grant No. 12535004, and USTC Research Funds of the Double First-Class Initiative No.\ YD2030002017 (China); the Ministry of Education, Youth and Sports (MEYS) of the Czech Republic; the Shota Rustaveli National Science Foundation, grant FR-22-985 (Georgia); the Deutsche Forschungsgemeinschaft (DFG), among others, under Germany's Excellence Strategy -- EXC 2121 ``Quantum Universe" -- 390833306, and under project number 400140256 - GRK2497; the Hellenic Foundation for Research and Innovation (HFRI), Project Number 2288 (Greece); the Hungarian Academy of Sciences, the New National Excellence Program - \'UNKP, the NKFIH research grants K 131991, K 133046, K 138136, K 143460, K 143477, K 146913, K 146914, K 147048, 2020-2.2.1-ED-2021-00181, TKP2021-NKTA-64, and 2025-1.1.5-NEMZ\_KI-2025-00004 (Hungary); the Council of Science and Industrial Research, India; ICSC -- National Research Center for High Performance Computing, Big Data and Quantum Computing, FAIR -- Future Artificial Intelligence Research, and CUP I53D23001070006 (Mission 4 Component 1), funded by the NextGenerationEU program (Italy); the Latvian Council of Science; the Ministry of Education and Science, project no. 2022/WK/14, and the National Science Center, contracts Opus 2021/41/B/ST2/01369, 2021/43/B/ST2/01552, 2023/49/B/ST2/03273, and the NAWA contract BPN/PPO/2021/1/00011 (Poland); the Funda\c{c}\~ao para a Ci\^encia e a Tecnologia (Portugal); the National Priorities Research Program by Qatar National Research Fund; MICIU/AEI/10.13039/501100011033, ERDF/EU, ``European Union NextGenerationEU/PRTR", and Programa Severo Ochoa del Principado de Asturias (Spain); the Chulalongkorn Academic into Its 2nd Century Project Advancement Project, the National Science, Research and Innovation Fund program IND\_FF\_68\_369\_2300\_097, and the Program Management Unit for Human Resources \& Institutional Development, Research and Innovation, grant B39G680009 (Thailand); the Eric \& Wendy Schmidt Fund for Strategic Innovation through the CERN Next Generation Triggers project under grant agreement number SIF-2023-004; the Kavli Foundation; the Nvidia Corporation; the SuperMicro Corporation; the Welch Foundation, contract C-1845; and the Weston Havens Foundation (USA).
\end{acknowledgments}

\bibliography{auto_generated}
\cleardoublepage \appendix\section{The CMS Collaboration \label{app:collab}}\begin{sloppypar}\hyphenpenalty=5000\widowpenalty=500\clubpenalty=5000
\cmsinstitute{Yerevan Physics Institute, Yerevan, Armenia}
{\tolerance=6000
A.~Hayrapetyan, V.~Makarenko\cmsorcid{0000-0002-8406-8605}, A.~Tumasyan\cmsAuthorMark{1}\cmsorcid{0009-0000-0684-6742}
\par}
\cmsinstitute{Institut f\"{u}r Hochenergiephysik, Vienna, Austria}
{\tolerance=6000
W.~Adam\cmsorcid{0000-0001-9099-4341}, L.~Benato\cmsorcid{0000-0001-5135-7489}, T.~Bergauer\cmsorcid{0000-0002-5786-0293}, M.~Dragicevic\cmsorcid{0000-0003-1967-6783}, P.S.~Hussain\cmsorcid{0000-0002-4825-5278}, M.~Jeitler\cmsAuthorMark{2}\cmsorcid{0000-0002-5141-9560}, N.~Krammer\cmsorcid{0000-0002-0548-0985}, A.~Li\cmsorcid{0000-0002-4547-116X}, D.~Liko\cmsorcid{0000-0002-3380-473X}, M.~Matthewman, J.~Schieck\cmsAuthorMark{2}\cmsorcid{0000-0002-1058-8093}, R.~Sch\"{o}fbeck\cmsAuthorMark{2}\cmsorcid{0000-0002-2332-8784}, M.~Shooshtari\cmsorcid{0009-0004-8882-4887}, M.~Sonawane\cmsorcid{0000-0003-0510-7010}, N.~Van~Den~Bossche\cmsorcid{0000-0003-2973-4991}, W.~Waltenberger\cmsorcid{0000-0002-6215-7228}, C.-E.~Wulz\cmsAuthorMark{2}\cmsorcid{0000-0001-9226-5812}
% ...
\par}
\vskip\cmsinstskip
\dag:~Deceased\\
$^{1}$Also at Yerevan State University, Yerevan, Armenia\\
$^{2}$Also at TU Wien, Vienna, Austria\\
% ...
\end{sloppypar}
\end{document}